\newcommand{\vr}{ {\bf r}} 
\newcommand{\vrp}{ {\bf r'}} 
\newcommand{\vrpp}{ {\bf r''}} 
\newcommand{\vk}{ {\bf k}} 
\newcommand{\vG}{ {\bf G}} 
\newcommand{\expv}[1]{\ensuremath{\langle #1  \rangle}}

\newcommand{\bracket}[3]{\ensuremath{\langle #1 | #2 | #3 \rangle}}
\newcommand{\ket}[1]{\ensuremath{| #1 \rangle}}

\newcommand{\refeq}[1]{(\ref{#1})} 
\newcommand{\tt}[1]{{\textt{#1}}} 

\documentclass[10pt]{patiopart}

\usepackage{graphicx}
\usepackage{dcolumn}
\usepackage{bm}
\usepackage{epsfig}
\usepackage{verbatim}
\usepackage{multirow}

\newcolumntype{d}[1]{D{.}{.}{#1}}
\eqnobysec 

\begin{document}

\title[Combining $GW$ calculations with
       exact-exchange density-functional theory]
       {Combining $GW$ calculations with
       exact-exchange density-functional theory: An analysis
       of valence-band photoemission for compound semiconductors}
\author{Patrick Rinke$^1$, 
        Abdallah Qteish$^{1,2}$,
        J\"org Neugebauer$^{1,3,4}$,
	Christoph Freysoldt$^1$ and 
	Matthias Scheffler$^1$}
\address{$^1$\ Fritz-Haber-Institut der Max-Planck-Gesellschaft, D-14195
               Berlin-Dahlem, Germany}
\address{$^2$\ Department of Physics, Yarmouk University, 21163-Irbid, Jordan}
\address{$^3$\ Department of Theoretical Physics, University of Paderborn, 
             D-33095 Paderborn, Germany}
\address{$^4$\ Max-Planck-Institut f\"ur Eisenforschung, Department of
               Computational Materials Design, D-40237 D\"usseldorf, Germany}
	     
\ead{rinke@fhi-berlin.mpg.de}

\begin{abstract}
We report quasiparticle-energy calculations of the electronic
bandstructure as measured by valence-band photoemission 
for selected II-VI compounds and group-III-nitrides. 
By applying $GW$ as perturbation to the ground state of the 
fictitious, non-interacting Kohn-Sham
electrons of density-functional theory (DFT)
we systematically study the electronic structure
of zinc-blende GaN, ZnO, ZnS and CdS.
Special emphasis is put on analysing 
the role played by the cation semicore $d$-electrons that
are explicitly included as valence electrons in our pseudopotential approach.
Unlike in the majority of previous $GW$ studies, which are almost
exlusively based on ground state calculations in the local-density approximation
(LDA), we combine $GW$ with
exact-exchange DFT calculations in the optimised-effective potential
approach (OEPx). This is a much more elaborate and computationally expensive 
approach. However,
we show that applying the OEPx approach leads to an improved description 
of the $d$-electron hybridisation compared to the LDA. 
Moreover we find that it
is essential to use OEPx
pseudopotentials in order to treat core-valence exchange consistently. 
Our OEPx based quasiparticle valence bandstructures are in good agreement with
available photoemission data in contrast to the ones based on the LDA.
We therefore conclude that for these materials OEPx constitutes
the better starting point for subsequent $GW$ calculations.

\end{abstract}

\submitto{\NJP}



\maketitle

\tableofcontents

\section{Introduction}
From the discovery of the photo-electric effect in the early days of the 20th
century photo-electron spectroscopy has developed into an invaluable 
experimental 
method for the study of electron excitations in bulk solids and surfaces.
Photoemission spectroscopy (PES) 
\cite{Himpsel:1983,Plummer/Eberhardt:1982,ARPES:1992}
and its inverse counterpart (IPES) \cite{Dose:1985,Smith:1988,IPES:1992}
have been instrumental for our current
understanding of elementary excitation processes in condensed matter and for
deciphering the electronic structure of many materials.
The success of PES and IPES ows much to the interpretation of the photo-electron
spectra in terms of single-particle excitations or \emph{quasiparticles} 
in the language of many-body quantum-mechanics.

In the first part of this article we will briefly recapitulate this connection 
between photo-electron spectroscopy of delocalised valence states and Green's 
function theory and illustrate how angular resolved
(I)PES spectra can be interpreted in terms of the quasiparticle bandstructure.
Within the theoretical framework of many-body perturbation theory we employ
Hedin's $GW$ approximation \cite{Hedin:1965} to 
calculate the quasiparticle energy spectrum, where $G$ refers to the Green's
function and $W$ to the dynamically screened Coulomb interaction. The $GW$ method
and the computational details of the bandstructure
calculations from the state-of-the-art (based on the
local-density-approximation) to recent developments (optimised effective
potential method to density-functional theory) will be introduced in more detail
later.
For further reading with regard to the connection between photo-electron
spectroscopy and many-body
perturbation theory we refer to the extensive review by 
Onida, Reining and Rubio \cite{Onida/Reining/Rubio:2002}.

In this article we report calculations of the quasiparticle
bandstructure of GaN and the II-VI compounds ZnO, ZnS and CdS in the zinc-blende
structure. The $GW$ method is defined as a perturbation to a system of 
non-interacting electrons, and we use density-functional theory (DFT) together
with the Kohn-Sham concept of \emph{fictitious} non-interacting particles as 
starting point for our calculations. 
Unlike in Kohn-Sham DFT  
a self-consistent solution of the many-body perturbation would successively
introduce higher order electron-electron interactions with every iteration step.
These enter in the set of Hedin's equation for the Green's function
\cite{Hedin:1965} through the vertex function $\Gamma$, which up until now can
only be solved fully for simple model systems.
For this reason we refrain from
any self-consistent treatment within the $GW$ equations 
themselves\footnote{ We will
briefly allude to some of the controversial issues pertaining to 
self-consistency in $GW$ in later sections.} and 
remain with the zeroth order in the self-energy ($\Sigma_0=iG_0W_0$), 
which typically gives bandstructures for weakly correlated quasiparticles in good
agreement with valence-band photo-electron spectroscopy 
\cite{Aulbur/Jonsson/Wilkins:2000,Aryasetiawan/Gunnarsson:1998}.
In the $G_0W_0$ approximation to $\Sigma$ the input 
Green's function and thus also the self-energy
becomes dependent on the ground state calculation and hence the
exchange-correlation functional used.
This dependence is a central aspect of this article. 

The Kohn-Sham eigenvalues of the 
commonly used local-density approximation (LDA) to the
exchange-correlation potential give a particularly poor account of the
electronic bandstructure in the II-VI materials and also to some extend in GaN. 
This is largely due to the inherent self-interaction effects in the LDA
introduced by the cation semicore $d$-electrons
\cite{Vogel/Krueger/Pollmann:1995,Vogel/Krueger/Pollmann:1996,
      Vogel/Krueger/Pollmann:1997}. 
The exact-exchange approach to
density-functional theory (EXX) on the other hand is naturally 
self-interaction free. Contrary to the LDA and its gradient-corrected flavours
(GGAs) the exchange potential in EXX-DFT only implicitly depends on
the electron density via the Kohn-Sham orbitals, a point that we will further 
elucidate in later sections.
Since the local exchange potential is constructed from 
the non-local Fock operator via the optimized effective potential method we
prefer to abbreviate this approach by OEPx instead of EXX.

Due to its relationship with Hartree-Fock, OEPx
has so far featured more 
prominently in atomic and molecular physics but recently applications to 
extended systems have also flourished
\cite{Kotani:1994,Kotani:1995,Kotani/Akai:1996,
      Goerling:1996,Staedele/etal:1997,Staedele/etal:1999,
      Aulbur/Staedele/Goerling:2000,Fleszar:2001,
      Gall/etal:2001,Fitzer/etal:2003,Magyar/Fleszar/Gross:2004,
      EXXpaper:2004}. 
In these calculations a remarkably good agreement with experiment for the band
gaps over a range of semiconductors has been reported. All
DFT exact-exchange studies for 
group-III-nitrides and II-VI compounds 
\cite{Staedele/etal:1997,Staedele/etal:1999,
      Aulbur/Staedele/Goerling:2000,Fleszar:2001}, however, have in common
that the cation $d$-electrons have effectively been removed from the calculation
by modelling their interaction with the valence electrons 
in terms of a pseudopotential. Although freezing the $d$-electrons in the core
of a pseudopotential is
computationally very efficient, it leads to a distinct disagreement between
theory and experiment for the strucural properties as shown by several LDA 
studies
\cite{Fiorentini/Methfessel/Scheffler:1993,Schroer/Krueger/Pollmann:1992,
      Schroer/Krueger/Pollmann:1993}. 
In all calculations presented in this article the $d$-electrons were explicitly
taken into account. Apart from improving the structural properties the 
inclusion of the
$d$-electrons also has a profound effect on the electronic structure.
We find that applying the OEPx approach leads to an improved 
description 
of the $d$-electron hybridisation compared to the LDA and therefore also to a
better agreement between (I)PES data and our quasiparticle energy calculations.

If applied in an all-electron fashion, $GW$
excited state calculations based on an LDA ground state will to a large extend
remove the spurious self-interaction inherent to the LDA   
\cite{Kotani/Schilfgaarde:2002,Usuda/Schilfgaarde:2002,Usuda/Hamada/etal:2004,
      Faleev/Schilfgaarde/Kotani:2004}. 
When a pseudopotential concept is applied on the other hand the self-interaction
of those core states that are locked away in the pseudopotential cannot be
corrected by the $GW$ approach. Thus, when semicore states that
contribute to the chemical bonding or hybridise with
valence states are present, all core states whose wavefunctions overlap strongly with 
these semicore states must be included explicitly      
as previous LDA based $GW$ calculations demonstrate
\cite{Rohlfing/Pollmann:1995,Rohlfing/Pollmann:1998,Luo/Louie:2002}.
A key result of or our approach is that if exact-exchange pseudopotentials 
\cite{Moukara/etal:2000,Engel/etal:2001} are used in the OEPx calculations
it is sufficient to treat only the $d$-electrons as valence electrons.
The $s$ and $p$ electrons of the same shell can
be frozen in the core of the pseudopotential,
because the self-interaction is already absent 
from the exact-exchange pseudopotentials and hence will not have to be removed
by the $GW$ calculation. The absence of self-interaction from the OEPx
ground state gives rise to the aformentioned improvement in the description 
of the $d$-electron hybridisation compared to the LDA. This leads us to the
hypothesis that the exact-exchange ground state
constitutes the better starting point for subsequent $GW$ calculations for this
material class.

In fact up until now only two exact-exchange based
quasiparticle energy calculations have been reported in the literature for GaN
\cite{Aulbur/Staedele/Goerling:2000,Fleszar:2001}. 
In both cases, however, the $d$-electrons were treated as part of 
the frozen core. In this article we report for the first time exact-exchange 
calculations including the $d$-electrons.
For II-VI compound our exact-exchange based $GW$ studies are the first so far.

\section{Probing the Electronic Structure by Photoemission}

\subsection{Photo-Electron Spectroscopy and the Quasiparticle Concept}
\label{Sec:PES}

\begin{figure}
  \begin{center}
    \includegraphics[width=0.8\columnwidth]{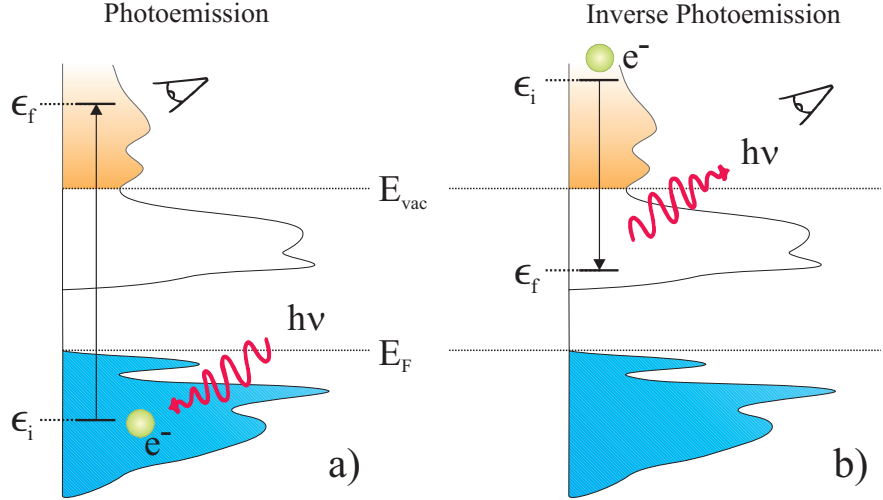}
    \caption{\label{fig:PES} 
             Schematic of the photoemission (PES) and inverse photoemission
	     (IPES) process. In PES (left) an electron is excited from an
	     occupied valence state (lower shaded region) into the 
	     continuum (upper shaded region) starting above the vacuum level
	     $E_{vac}$  by an
	     incoming photon. In IPES (right) an injected electron with kinetic
	     energy $\epsilon_i=E_{kin}$ undergoes a radiative transition into
	     an unoccupied state (white region) emitting a photon in the
	     process.}
  \end{center}
\end{figure}

In photo-electron spectroscopy (PES) electrons are ejected from a sample
upon irradiation with visible or ultraviolett light (UPS)
or with X-rays (XPS), as sketched in Fig.  \ref{fig:PES}a.
The energy of the bound electron states $\epsilon_i$ can be reconstructed
from the photon energy $h\nu$ and
the kinetic energy $E_{kin}$ of the photoelectrons that reach the 
detector\footnote{Throughout this article the top of the valence bands is 
chosen as energy zero.}
\begin{equation}
\label{Eq:def:Ei}
  \epsilon_i= h \nu - E_{kin}\quad.
\end{equation}
Equation \refeq{Eq:def:Ei} defines the binding energy of the electron in the
solid. 

By inverting  the photoemission process, as schematically shown in 
Fig. \ref{fig:PES}b, the unoccupied states can be probed. 
An incident electron with energy $E_{kin}$ is scattered in the
sample emitting \emph{bremsstrahlung}. Eventually it will undergo
a radiative transition into a lower-lying unoccupied state, emitting a photon
that carries the transition energy $h \nu$. 
The energy of the final, unoccupied state can
be deduced from the measured photon energy according to
\begin{equation}
\label{Eq:def:Ef}
  \epsilon_f = E_{kin} - h \nu \quad.
\end{equation}
This technique is commonly referred to as inverse photoemission spectroscopy
(IPES) or bremsstrahlung isochromat spectroscopy (BIS).

The experimental observable in photoemission spectroscopy is the photocurrent.
Since the energy dependence
of the transition matrix elements is usually weak and smooth,
structures in the photoemission spectrum can be associated with
features in the density of states (DOS), i.e. the imaginary
part of the one-particle Green function
\footnote{Atomic units 
$4\pi\epsilon_0=h=e=m_e=1$, where $e$ and $m_e$ are the charge and mass of an
electron, respectively, will be used in the remainder of this article.} 
\cite{Onida/Reining/Rubio:2002,Almbladh/Hedin:1983}
\begin{equation}
\label{Eq:def_A}
  A(\vr,\vrp;\epsilon)=\frac{1}{\pi}{\rm Im} \: G(\vr,\vrp;\epsilon)
                      =\sum_s
		       \psi_s(\vr)\psi_s^*(\vrp)
		       \delta(\epsilon-\epsilon_s) \quad .
\end{equation}
The sum in the last expression runs over all states $s$ the system can 
assume
and the photocurrent is then the surface weighted integral over the diagonal
part of the spectral function $A(\vr,\vrp;\epsilon)$. We note, however, that
with respect to the measured intensities a
photoemission spectrum should be viewed as a noticeably distorted
spectral function. In particular when selection rules become important certain
peaks in the spectral function may be significantly reduced or may even 
disappear completely. 
The energies $\epsilon_s$ in equation \refeq{Eq:def_A} are the excitation 
energies of the many-body state
created by the addition or removal of the photo-electron and
$\psi_s(\vr)$ gives the transition amplitude from the $N$ to the
$N\pm 1$-body state. In many-body quantum mechanics $\epsilon_s$ and
$\psi_s(\vr)$ are defined  as
\cite{ARPES:1992,IPES:1992,Almbladh/Hedin:1983,Bardyszewski/Hedin:1985,
      Inglesfield:1987,Hedin/Lundqvist:GW,Gross/Runge:MPT}: 
\begin{eqnarray}
\label{Eq:def_es}
  \left.
  \begin{array}{lcl}
    \epsilon_s&=&E(N+1,s)-E(N) \\
    \psi_s(\vr)&=&\bracket{N}{\hat{\psi}(\vr)}{N+1,s}
  \end{array}
  \quad \right\} \quad 
  \textrm{for} \quad \epsilon_s \geq E_F \\
\label{Eq:eng_ex_h}
  \left.
  \begin{array}{lcl}
    \epsilon_s&=&E(N)-E(N-1,s)\\
    \psi_s(\vr)&=&\bracket{N-1,s}{\hat{\psi}(\vr)}{N}
  \end{array}
  \quad \right\} \quad 
  \textrm{for} \quad \epsilon_s < E_F
\end{eqnarray}
Here $E_F$ is the Fermi energy of the system. 
The states $\ket{N,s}$ are many-body eigenstates of the
$N$-electron Schr\"odinger equation $\hat{H}\ket{N,s}=E(N,s)\ket{N,s}$,
$\hat{H}$ is the many-body Hamiltonian and 
$E(N,s)=\bracket{N,s}{\hat{H}}{N,s}$ is the corresponding total energy. 
The field operator $\hat{\psi}(\vr)$ annihilates an electron from the 
many-body states $\ket{N+1}$ or $\ket{N}$. 
The representation given in equation \refeq{Eq:def_es} and
\refeq{Eq:eng_ex_h} is particularly insightful because it allows a
direct interpretation of $\epsilon_s$ 
as photo-excitation energy 
from the $N$-particle ground state with total energy $E(N)$ into an excited 
state $s$ of the ($N$-1)-particle system with total energy $E(N-1,s)$ 
upon removal of
an electron in the photoemission process. Similarly the addition energy that is
released in the radiative transition in
inverse photoemission is given by the total energy difference of the 
excited ($N$+1)-particle system and the ground state.

Via the field operator  formalism the many-body Hamiltonian
can be transformed into a single-particle 
Hamiltonian\footnote{Since the nuclei are assumed to be stationary 
throughout this article, 
the nucleus-nucleus interaction contributes only a constant to the total
energy, whereas the electron-nucleus interaction can be represented by an 
external potential $v_{ext}(\vr)$.} \cite{Gross/Runge:MPT}:
$\hat{H}(\vr,\vrp;\epsilon)=\hat{h_0}(\vr)+\Sigma(\vr,\vrp;\epsilon)$.
All electron-electron
interaction terms are rolled up in the non-local, energy dependent self-energy
$\Sigma$ and the remaining contributions are given by
$\hat{h}_0(\vr)=-\frac{1}{2} \nabla^2+v_{ext}(\vr)$.  
The single particle Green's function 
\begin{equation}
\label{Eq:G_Leh}
  G(\vr,\vrp;\epsilon)=\bracket{\vr} 
                          {[\hat{H}(\epsilon)-\epsilon]^{-1}}{\vrp}
                      =\lim_{\eta \rightarrow 0^+}
                           \sum_s \frac{\psi_s(\vr) \psi_s^*(\vrp)}
                           {\epsilon-(\epsilon_s 
			      + i\eta\:{\rm sgn}(E_f-\epsilon_s))}
\end{equation}
then satisfies the Dyson equation 
\begin{equation}
\label{Eq:G_eq}
  G^{-1}(\vr,\vrp;\epsilon) = \left[\epsilon-\hat{h}_0(\vr) \right] \delta(\vr-\vrp)
                            - \Sigma(\vr,\vrp;\epsilon) 
\end{equation}
and by inserting equation \refeq{Eq:G_Leh} into \refeq{Eq:G_eq} one 
immediately finds that $\epsilon_s$ and $\psi_s(\vr)$  are solutions to
\begin{equation}
\label{Eq:Dyson}
  \hat{h}_0(\vr)\psi_s(\vr)+ \int \! d\vrp 
          \Sigma(\vr,\vrp;\epsilon_s)  \psi_s(\vrp) = \epsilon_s \psi_s(\vr)
	  \quad .
\end{equation}
The poles of the Green's function therefore
correspond to the real electron addition and removal
energies $\epsilon_s$ and form a branch-cut infinitesimally above (below) 
the real energy axis for occupied (unoccupied) states.

So far \refeq{Eq:def_es} to \refeq{Eq:Dyson} have been
exact, which is of limited use for practical computational schemes. 
To establish a link to photoemission
spectroscopy of delocalised valence states it is helpful to introduce Landau's
concept of \emph{quasiparticles} \cite{Landau}. 
This new entity can be considered as a
combination of an electron or hole with its surrounding polarisation cloud or in
other words as
the collective response of the interacting many-body system upon 
photo-excitation. Switching to the quasiparticle picture is consistent with
analytically continuing the self-energy to the complex energy domain. The
quasiparticle poles (now at complex energy) each represent the effect of 
many of the infinitesimally closely spaced poles just above (below) the real
axis. We will return to this point later in this section.

To motivate the association of \emph{quasiparticles} with particle-like
excitations we turn again to the spectral function. For non-interacting
electrons the spectral function consists of a series of delta peaks
\begin{equation}
  A_{sn}(\epsilon)=\bracket{\psi_s(\vr)}{A(\vr,\vrp;\epsilon)}{\psi_n(\vrp)}=
                   \delta_{sn}\delta(\epsilon-\epsilon_s), 
\end{equation} 
each of which corresponds to the excitation of a particle.
The many-body states $\ket{N}$
and $\ket{N\pm 1}$ are represented by a single Slater determinant and the
excitated state configurations by a single creation ($\hat{a}_s$) or 
anihilation ($\hat{a}^\dagger_s$) operator acting on the ground state: 
$\ket{N+1,s}=\hat{a}^\dagger_s\ket{N}$.
The excitation energies $\epsilon_s$ and the wavefunctions $\psi_s(\vr)$ are
thus the
eigenvalues and eigenfunctions of the single-particle Hamiltonian.

When the electron-electron interaction is turned on, the electrons can no 
longer be regarded as independent particles. As a consequence 
the matrix elements of the spectral function $A_{sn}(\epsilon)$ will
contain contributions from many non-vanishing transition amplitudes. 
If these contributions merge into a clearly identifiable peak 
that appears to be derived from a single delta-peak broadened by the 
electron-electron interaction this
structure can be interpreted as single-particle like excitation -- the
\emph{quasiparticle}. The broadening of the quasiparticle peak in the spectral
function is associated with the lifetime $\tau$ of the excitation due to
electron-electron scattering, whereas
the area underneath the peak is interpreted
as the renormalisation $Z$ of the quasiparticle. 
This renormalisation factor quantifies the reduction in spectral
weight due to electron-electron exchange and correlation effects 
compared to an independent electron. In summary a quasiparticle
peak in the spectral function will exhibit the following shape
\begin{equation}
  A_s(\epsilon)\approx\frac{Z_s}{\epsilon-(\epsilon_s+i\Gamma)}
  \quad.
\end{equation}
In contrast to the exact energies of the many-body states, 
which are poles of the Green's function
on the real axis, the quasiparticle poles reside in the complex plane and are no
longer eigenvalues of the single-particle Hamiltonian.
The real part of this complex energy is associated with the
energy of the quasiparticle excitation and the imaginary part with its 
inverse lifetime $\Gamma=2/\tau$. 

Apart from quasiparticle excitations a typical photoemission experiment 
provides a rich variety of additional information. 
In core-electron emission for
instance inelastic losses or multi-electron excitations such as shake-ups and 
shake-offs lead to satellites in the spectrum. These are genuine 
many-body effects that are not contained in the quasiparticle approximation.
However, since we are primarily interested in the description of valence 
bands in 
semiconductors these many-body effects are not important for interpreting
the spectral features. For a more indepth discussion of these many-body effects
we therefore refer the interested reader to the following articles 
\cite{Almbladh/Hedin:1983,Gadzuk:1978,Lee/Gunnarsson/Hedin:1999,Hedin:1999,
      Rehr/Albers:2000}.

Before we introduce the $GW$ approximation to the self-energy as 
a tractable computational approach for calculating the quasiparticle energies
we will briefly address the reconstruction of bandstructure information from the
measured photo-electron spectra.
By varying the angle of incidence
(angle resolved  photo-electron
spectroscopy (ARPES) \cite{ARPES:1992} and $\vk$-resolved 
inverse photo-electron spectroscopy
(KRIPES) \cite{IPES:1992}) dispersion relations of the 
excitated states can be obtained. However,
since the emitted photons or electrons inevitably have to pass the surface of
the crystal to reach the detector, information about their transverse
momentum $k_\bot$ is lost. This is due to the fact that the
translational invariance  is broken at the surface and
only the parallel momentum component $k_{\parallel}$ is conserved.

In order to reconstruct the three-dimensional bandstructure of the solid 
assumptions are often made about the dispersion of the final states 
\cite{Himpsel:1983,Plummer/Eberhardt:1982,Dose:1985,
      Smith:1988,Hora/Scheffler:1984}. Alternatively
{\it ab initio} calculations as described in this article can aid in the
assignment of the measured peaks. 
Only recently absolute band mapping has been reported in
secondary electron emission (SEE) experiments \cite{Bovet/etal:2004} 
and also for a technique combining 
ARPES with very low-energy electron diffraction (VLEED) 
\cite{Strocov/etal:1998}.

State of the art spectroscopy techniques of course 
allow the variation of many 
more parameters than just the angle of incidence. A more in-depth discussion 
of photoemission experiments will, however, go beyond the scope of this 
article and we refer the reader to 
references \cite{ARPES:1992} and \cite{IPES:1992} for more details.

\subsection{The $GW$ Formalism}
\label{Sec:GW}

To solve the Dyson equation
\refeq{Eq:G_eq} for real systems 
one typically applies Hedin's $GW$ approximation \cite{Hedin:1965} for the
self-energy.
Assuming that the quasiparticles interact only weakly via the screened Coulomb
interaction $W$ the self-energy in $GW$ is then given as
\begin{equation}
\label{Eq:S=GW}
  \Sigma_{xc}^{GW}(\vr,\vrp;\epsilon)=\frac{i}{2\pi}\int_{-\infty}^\infty \! d\epsilon' 
                          e^{i\epsilon'\delta}
                          G(\vr,\vrp;\epsilon+\epsilon')W(\vr,\vrp;\epsilon')
\end{equation}
where $\delta$ is an infinitesimal, positive time.
In practice one starts from a system of non-interacting particles with energies
$\epsilon_i$ and wavefunctions $\phi_i(\vr)$. The non-interacting Green's 
function is defined analogous to equation \refeq{Eq:G_Leh} as\footnote{Note that only
spin unpolarised systems are considered here. All state summations therefore
include the spin variable implicitly.}
\begin{equation}
\label{Eq:G_0}
  G_0(\vr,\vrp,\epsilon)= \lim_{\eta \rightarrow 0^+} \sum_i 
                  \frac{\phi_i(\vr) \phi_i^*(\vrp)}
                  {\epsilon-(\epsilon_i + i\eta\:{\rm
		  sgn}(E_f-\epsilon_i))}. 
\end{equation}
The quantum state indices $i$ and $s$ 
are short for the composite of band index $n$ and wave vector $\vk$. 
In the random phase approximation (RPA) the dielectric function 
\begin{equation}
\label{Eq:eps_RPA}
  \varepsilon(\vr,\vrp,\epsilon)=  \delta(\vr-\vrp) -
                             \int \! d\vrpp v(\vr-\vrpp)
			     \chi_0(\vrpp,\vrp;\epsilon)
\end{equation}
is connected to the independent particle polarisability
\begin{equation}
\label{Eq:chi_0}
  \chi_0(\vr,\vrp,\epsilon)= -\frac{i}{2\pi}\int_{-\infty}^\infty \! d\epsilon' 
                                G_0(\vr,\vrp;\epsilon'-\epsilon)
			        G_0(\vrp,\vr;\epsilon')
\end{equation}
and the bare Coulomb interaction
\begin{equation}
\label{Eq:v}
  v(\vr-\vrp)= \frac{1}{|\vr-\vrp|}
\end{equation}
is screened by the inverse dielectric function 
\begin{equation}
\label{Eq:W}
  W_0(\vr,\vrp,\epsilon)= \int \! d\vrpp \varepsilon^{-1}(\vr,\vrpp;\epsilon) 
                                   v(\vrpp-\vrp) \quad.		      
\end{equation}
For numerical convenience and physical insight we separate the $GW$ self-energy 
\refeq{Eq:S=GW} according to $\Sigma_{xc}^{GW}=\Sigma_x+\Sigma_c^{GW}$ 
with the two terms defined as
\begin{eqnarray}
\label{Eq:def:sigma_x}
  \Sigma_x(\vr,\vrp)     &=\frac{i}{2\pi}\int_{-\infty}^\infty \! d\epsilon'
                           e^{i\epsilon'\delta}
			   G(\vr,\vrp;\epsilon')v(\vr-\vrp) \\ 
			 & =-\sum_{i}^{occ} 
                      \phi_i(\vr)v(\vr-\vrp)\phi_i^*(\vrp) \nonumber \\ 
\label{Eq:def:sigma_c}
  \Sigma_c^{GW}(\vr,\vrp;\epsilon)&=\frac{i}{2\pi}\int_{-\infty}^\infty \! d\epsilon'
                    e^{i\epsilon'\delta} G(\vr,\vrp;\epsilon+\epsilon')
                    \left[W(\vr,\vrp;\epsilon')-v(\vr-\vrp)\right]
\end{eqnarray}
$\Sigma_x$ is the Fock or bare exchange operator that we will revisit in section
\ref{Sec:OEP} and $\Sigma_c^{GW}$
encompasses the dynamic correlation of the quasiparticles. Note that the
definition of the self-energy in equation \refeq{Eq:G_eq} implicitly
includes the Hartree potential
\begin{equation}
\label{Eq:def:vHartree}
  v_H(\vr)=\int \! d\vrp \: n(\vrp) v(\vr-\vrp)
\end{equation}
with $n(\vr)$ being the electron density, whereas in this section the Hartree
potential is 
separated from the $GW$ self-energy. 
Inserting $\Sigma_{xc}$ and the quasiparticle Green's function \refeq{Eq:G_Leh}
into \refeq{Eq:G_eq} the Dyson equation becomes
\begin{equation}
\label{Eq:qp_eq}
  \left[\hat{h}_0(\vr)+v_H(\vr)\right] \psi_s(\vr) +
  \int \! d\vrp \Sigma_{xc}^{GW}(\vr,\vrp;\epsilon_s) \psi_s(\vrp) = 
  \epsilon_s \psi_s(\vr)
\end{equation}
This equation, also referred to as \emph{quasiparticle equation}, 
can then be solved for the quasiparticle energies and wavefunctions.

In principle
the set of equations \refeq{Eq:S=GW} to \refeq{Eq:W} could be solved
self-consistently via the use of the Dyson equation \refeq{Eq:G_eq} 
expressed now in terms of the non-interacting Greens function $G_0$ 
(and for brevity written in operator notation)
\begin{equation}
\label{Eq:Dyson_eq}
  G=G_0+G_0\left[v_H+\Sigma_{xc}^{GW}\right]G \quad .
\end{equation}
A crucial point to note, however, is that at each iteration higher order
diagrams would have to be included, because the $GW$ approximation is only 
equivalent to the first iteration of Hedin's equations
\cite{Hedin:1965}, which are an exact set of
equations for the Green's function and the self-energy. Both the
polarisability $P=-iGG\Gamma$ and the self-energy $\Sigma=iG\Gamma W$ contain the
vertex function $\Gamma$. Solving equations \refeq{Eq:S=GW} to 
\refeq{Eq:Dyson_eq} \emph{self-consistenly} is therefore \emph{inconsistent}
with Hedin's equations if no higher order electron-electron interactions are
included via the vertex function $\Gamma$.

We can nevertheless group  solutions to Hedin's $GW$ equations \refeq{Eq:S=GW} to \refeq{Eq:W}
into three  categories: 
self-consistent ($\Sigma=iGW$), partially self-consistent ($\Sigma=iGW_0$) and 
non self-consistent ($\Sigma=iG_0W_0$). 
Only $GW_0$ and $GW$ fulfill certain sum rules including 
particle number conservation \cite{Schindlmayr/Garcia-Gonzalez/Godby:2001}, which
give rise to an improved description of ground state total energies
\cite{Holm/vonBarth:1998,Garcia-Gonzalez/Godby:2001}.
Spectral features, on the other hand, broaden
with increasing number of iterations in the self-consistenty cycle, as was first
observed for the homogeneous electron gas \cite{Holm/vonBarth:1998}. Similarly, 
for closed shell atoms the good 
agreement with experiment for the ionisation potential in $G_0W_0$ is 
lost upon iterating the equations to self-consistency
\cite{Delaney/Garcia-Gonzalez/Rubio/Rinke/Godby:2004}. 

For bulk materials  self-consistent $GW$ calculations also exhibit a 
broadening of the spectral features compared to $G_0W_0$
\cite{Schoene/Eguiluz:1998,Tamme/Schepe/Henneberger:1999,
      Schoene/Eguiluz:1999,Ku/Eguiluz:2002}.
In pseudopotential $GW$ calculations for bulk silicon this leads to a gross 
overestimation of the fundamental band gap \cite{Schoene/Eguiluz:1998}, 
whereas an all-electron $GW$ calculations yield band gaps in seemingly good
agreement with experiment \cite{Ku/Eguiluz:2002}. This discrepancy between
pseudopotential and all-electron $GW$ calculations has, to the best of our
knowledge, not yet been resolved and is currently being discussed controversally
in the literature (see for instance 
\cite{Faleev/Schilfgaarde/Kotani:2004,
      Delaney/Garcia-Gonzalez/Rubio/Rinke/Godby:2004,
      Tiago/Ismail-Beigi/Louie:2003}).

Since addressing the issue of self-consistency in more detail would 
lead beyond the scope of this article
we terminate the self-consistency cycle in our quasiparticle energy calculations
after the first iteration,
when the self-energy is given by $\Sigma_0^{GW}=iG_0W_0$ and solve the quasiparticle
equation \refeq{Eq:qp_eq} for the excitation energies.
This procedure implies, however, that the quasiparticle spectrum might now
depend on the input Green's function, $G_0$, a crucial aspect that we will
address in the following.

\subsection{DFT and the Kohn-Sham Bandstructure}
\label{Sec:DFT}
Density-functional theory (DFT) is probably the most widely used computational 
electronic structure method today for systems containing a large number of atoms.
The central quantities in DFT are the 
electron density $n(\vr)$ and the total energy $E_{tot}$. 
The latter is a functional of the former
and attains its minimum at the exact ground state density, as proven by Hohenberg and Kohn
\cite{Hohenberg/Kohn:1964}. This formalism was turned into a tractable
computational scheme by Kohn and Sham \cite{Kohn/Sham:1965}, by observing that
the system of interacting particles can be mapped onto a fictious system of
non-interacting particles that reproduce the same density as the many-body
problem of interest. 

Kohn and Sham divided the total energy into known contributions such as the 
kinetic energy of the non-interacting particles $T_s$, the
Hartree energy
\begin{equation}
\label{Eq:EH}
  E_{H}[n]=\frac{1}{2}\int \! d\vr \: n(\vr) v_{H}(\vr)
\end{equation}
and the external energy
\begin{equation}
\label{Eq:Eext}
  E_{ext}[n]=\int \! d\vr \: n(\vr) v_{ext}(\vr) \quad.
\end{equation}
and an unknown remainder. This last term includes all electron-electron 
interactions beyond the Hartree 
mean-field and is defined as the exchange-correlation energy 
\begin{equation}
\label{Eq:Exc}
  E_{xc}[n]=E_{tot}[n]- T_s[n]-E_{ext}[n]-E_H[n] \quad.
\end{equation}
The electron density
\begin{equation}
\label{Eq:KS_dens}
  n(\vr) = \sum_i^{occ} |\phi_i(\vr)|^2
\end{equation}
is composed of the occupied Kohn-Sham orbitals ${\phi_i(\vr)}$ that are
solutions of the Kohn-Sham equation
\begin{equation}
\label{Eq:KS_eq}
  \left[-\frac{\nabla^2}{2}+v_{ext}(\vr)+v_H(\vr) + v_{xc}(\vr) \right] \phi_i(\vr)  = 
          \epsilon_i \phi_i(\vr) \quad.
\end{equation}
The exchange-correlation potential $v_{xc}(\vr)$ 
is formally defined as the functional derivative of the 
exchange-correlation energy
\begin{equation}
\label{Eq:def:v_xc}
  v_{xc}(\vr)=\frac{\delta E_{xc}[n]}{\delta n(\vr)} \quad.
\end{equation}
Equations \refeq{Eq:EH} to \refeq{Eq:KS_eq} have to be solved
self-consistently until convergence in the total energy is reached. 

Since the exact form of the exchange-correlation functional
is unknown\footnote{To be more precise the exact dependence of $v_{xc}$
on the density alone is unknow. In the context of many-body perturbation
theory the exact exchange-correlation potential can be expressed in terms of the
Green's function and the self-energy via the Sham-Schl\"uter equation,
introduced in the next section. Alternatively an exact representation of
$v_{xc}$ can be obtained in G\"orling-Levy perturbation theory 
\cite{Goerling/Levy:PBT}. } 
suitable approximations have to be found in practice. In this
article we work in the local-density approximation (LDA) \cite{Kohn/Sham:1965}
or in the exact-exchange approximation to the optimised effective potential. 
Since the latter constitutes an important aspect of our work
we will describe it in more detail in the following sections.

Contrary to the poles of the Green's function \refeq{Eq:def_es} 
the Kohn-Sham eigenvalues are Lagrange multipliers and are therefore primarily
mathematical constructs.
Strictly spoken 
only the highest occupied Kohn-Sham eigenvalue of exact DFT can be 
rigorously assigned to the
ionisation potential \cite{Almbladh/Barth:1985,Levy/Perdew/Sahni:1984,
Casida:1999}. For an extended system with well defined chemical potential 
this is equivalent to stating that the chemical potential 
in DFT is the same as the true one (Janak's theorem \cite{Janak:1978}).
Furthermore Janak's theorem implies that for delocalised states the Kohn-Sham
eigenvalues are close to $\epsilon_i$ and $\epsilon_f$ as defined in equation
\refeq{Eq:def:Ei} and \refeq{Eq:def:Ef}. Recently further
justification for the
interpretation of exact Kohn-Sham orbital energies as approximate vertical 
ionization potentials was given for finite systems \cite{KS_I}. For
atoms and small molecules, where accurate {\it ab initio} densities were 
available, the Kohn-Sham eigenvalues were found to be very close to experimentally
measured ionisation and excitation 
energies \cite{KS_I,Savin/Umrigar/Gonze:1998}. Although,
noteable deviations were observed if an LDA or GGA functional was employed 
instead, the Kohn-Sham eigenvalues still provide a good starting point for 
self-energy calculations in the framework of many-body perturbation theory.

Furthermore the true $v_{xc}$ is a discontinuous function of the particle
number $N$, which essentially implies that the value of $v_{xc}$ jumps by a
constant for finite $\vr$ when the particle number is infinitesimally increased
from $N-\delta$ to $N+\delta$. This discontinuity has a profound effect on the 
calculated band gaps of semiconductors and insulators
\cite{Perdew/Levy:1983,Sham/Schlueter:1983}. 
In the LDA or GGA  the exchange-correlation potential
is a smoothly varying function with respect to changes in the particle number, 
whereas the exchange potential in the
exact-exchange formalism exhibits an integer discontinuity 
\cite{Krieger/Li/Iafrate:1992}, but not necessarily of the correct size.  
We will further allude to this point in the following sections.

\subsection{Connection between DFT and $GW$}
\label{Sec:SchSchl}

Before we introduce the exact-exchange approximation to $E_{xc}$ it is
elucidating to draw a connection between many-body perturbation theory as
described in section \ref{Sec:PES} and \ref{Sec:GW} and density-functional theory.
In the original proof \cite{Sham/Schlueter:1983,Sham/Schlueter:1985} 
Sham and Schl\"uter made use of the fact that the
Kohn-Sham density reproduces the exact electron density
(Hohenberg-Kohn theorem \cite{Hohenberg/Kohn:1964}). Both the density of the
interacting as well as the density of the non-interacting system 
can be obtained from the respective Green's
function: $n(\vr)=\frac{1}{\pi}\textrm{Im}\!\int \! d\epsilon 
G(\vr,\vr;\epsilon)$, which leads to the 
{\it density condition}
\begin{equation}
\label{Eq:densitycondition}
  0=n(\vr)-n_{\rm KS}(\vr)=
  \frac{1}{2\pi}{\rm Im} \! \int d\epsilon  \left[
	    G(\vr,\vr;\epsilon)-G_{\rm KS}(\vr,\vr;\epsilon)
	    \right].
\end{equation}
The Kohn-Sham Green's function, $G_{\rm KS}$, entering this equation is an 
independent particle Green's
function \refeq{Eq:G_0} constructed from the Kohn-Sham orbitals. 
Applying the Dyson equation \refeq{Eq:G_eq} transforms the density condition 
into the Sham-Schl\"uter equation
\begin{equation}
\label{Eq:def:ShScheq}
  \int d\epsilon \! \int \! d\vrp \! \int \! d\vrpp G_{\rm KS}(\vr,\vrp;\epsilon)
            \tilde{\Sigma}_{xc}(\vrp,\vrpp;\epsilon)
	    G(\vrpp,\vr;\epsilon)=0
\end{equation}
where the self-energy that connects the interacting with the non-interacting
system has been defined as
\begin{equation}
\label{Eq:def:diffSig}
  \tilde{\Sigma}_{xc}(\vr,\vrp;\epsilon)=\Sigma_{xc}(\vr,\vrp;\epsilon)-v_{xc}(\vr)
                            \delta(\vr-\vrp).
\end{equation}
The exchange-correlation potential of density-functional theory can therefore be
interpreted as the variationally best local approximation to the non-local,
dynamic self-energy \cite{Casida:1995}. With regards to the interpretation of 
the Kohn-Sham eigenvalues as photoemission excitation energies, however, 
the Sham-Schl\"uter equation corroborates the conjecture that one
has to go beyond the locality in time  or space  
to improve on the density-functional treatment, even if the 
exact $v_{xc}$ was known \cite{Casida:1995}.

\subsection{The Optimised Effective Potential Method and Exact-Exchange}
\label{Sec:OEP}
In this section we will introduce the exact-exchange (EXX) 
approximation to density-functional theory. Following Kohn and Sham's idea of
dividing the total energy into known and unknown contributions the
exact-exchange energy $E_x$ 
\begin{equation}
\label{Eq:EXX_Eng}
  E_x=-\frac{1}{2}\sum_{ij}^{occ}\int d\vr \int d\vrp
       \phi_i^*(\vr)\phi_j(\vr)
       v(\vr-\vrp)\phi_j^*(\vrp)\phi_i(\vrp)
\end{equation}
can be isolated from $E_{xc}$ leaving only the correlation part $E_{c}$ to be
approximated.
In the exact-exchange only approach this correlation term is ignored\footnote{
Later in this section we will reintroduce the correlation energy in an
approximate form that is commonly used in connection with exact-exchange DFT
calculations.} so that the 
total energy becomes
\begin{equation}
\label{Eq:KS_Etot_OEPx}
  E_{tot}^{\rm EXX} = T_s[n]+E_{ext}+E_H+E_{x} \quad .
\end{equation}
In order to connect to the previous section we will take a different route 
than in most texts to derive the Kohn-Sham exchange-potential from this energy 
expression. We will show  that  the Sham-Schl\"uter equation 
naturally reverts to the \emph{optimised effective potential} method (OPM) for
the exchange-correlation potential
\cite{Casida:1995,Grabo/Kreilich/Kurth/Gross:2000,Niquet/Fuchs/Gonze:2003,
      Engel:2003}.
We therefore chose to denote the optimised effective potential in 
the exact-exchange
approach by OEPx instead of EXX.

To derive the optimised effective potential (OEP) equations 
the Sham-Schl\"uter equation \refeq{Eq:def:ShScheq} is first 
linearised \cite{Godby/Schlueter/Sham:1986,Godby/Schlueter/Sham:1988,Casida:1995}
by replacing the fully interacting Green's function, $G$, with the 
Kohn-Sham Green's function, $G_{\rm KS}$.
Further replacing the self-energy with only the exchange part
\refeq{Eq:def:sigma_x}, which is
is equivalent to the Fock operator, and rearranging the
resulting equation
into the conventional form of a non-linear integral equation yields the
equation for the OEPx potential
\cite{Sham/Schlueter:1983,Grabo/Kreilich/Kurth/Gross:2000,
Engel:2003,Sharp/Horton:1953,Talman/Shadwick:1976}
\begin{equation}
\label{Eq:OEPx}
   \int \! d\vrp \chi_0(\vr,\vrp)v_{x}(\vrp)=\Lambda_{\rm OEPx}(\vr) 
\end{equation}
with 
\begin{equation}   
\label{Eq:def:Lambda_OEPx}
   \Lambda_{\rm OEPx}(\vr)=\! \int \! d\epsilon \! \int \! d\vrp \! \int \! 
                           d\vrpp 
                   G_{\rm KS}(\vr,\vrp;\epsilon)\Sigma_{x}(\vrp,\vrpp)
		   G_{\rm KS}(\vrpp,\vr;\epsilon). 
\end{equation}
$\chi_0(\vr,\vrp)$ is the independent particle polarisability,
$\chi_0(\vr,\vrp;\epsilon=0)$, as previously 
defined in the context of the $GW$ approximation (equation
\refeq{Eq:eps_RPA}).

The exchange-potential $v_x$ can be thought of
as the best local potential approximating the non-local Fock operator 
\cite{Casida:1995}. 
It is important to emphasise, however, that by construction 
the total energy in Hartree-Fock
is always smaller (or at most equal) and thus better than in the OEPx formalism
\cite{Ivanov/Levy:2003}, 
because the energy minimisation in the optimised effective potential method
is subject to the constraint of the wavefunctions being solutions to the
Kohn-Sham equation \refeq{Eq:KS_eq}. 
The eigenvalues of the
OEPx formalism, on the other hand, derive from an effective mean-field approach
for non-interacting particles, whereas
in Hartree-Fock they correspond to the energies of electrons interacting via
Pauli but not Coulomb correlation. The Kohn-Sham particles in the valence and 
conduction bands are therefore governed by the same effective potential, which
exhibits the correct asymptotic behaviour ($1/r$ decay for large distances in
finite systems) in the OEPx. 

In Hartree-Fock on the other hand 
virtual conduction electrons are only poorly accounted for and experience a 
different potential than the valence electrons. Since the Fock-operator contains
the self-interaction correction only for the valence electrons, 
a virtual conduction
electron interacts with all $N$ valence electrons, in contrast to a valence
electron, which interacts only with the remaining $N$-$1$ electrons. 
Thus, for a charge neutral 
system each valence electron in Hartree-Fock experiences a density of
netcharge +1 and the correct $1/r$ potential decay results for finite systems.
This stands in obvious contrast to the description of the virtual conduction
electrons, that see a neutral charge density. Since excitations in the
valence and conduction bands of solids leave the charge density largely 
unchanged the different treatment of valence and conduction electrons within
Hartree-Fock appears to be unjustified. Indeed,
as we will present in the following
sections the OEPx eigenvalues are closer to the
photo-electron excitation energies for the four semiconductors discussed in 
this article than the Hartree-Fock energies.

The exact-exchange potential $v_x$ is only implicitly a function of the
electron density via the Kohn-Sham orbitals. This is in accordance with the
Hohenberg-Kohn theorem as can easily be seen by performing the functional
derivative of the exact-exchange energy \refeq{Eq:EXX_Eng} with respect to the
density. Applying first order perturbation theory yields the 
more familiar expression for the OEPx
equation\footnote{Likewise this equation can be obtained from \refeq{Eq:OEPx} by integrating out
the frequency dependence of the inhomogeneity $\Lambda_{\rm OEPx}$, 
which can be done
analytically because $\Sigma_x$ is a static operator.
}
\begin{equation}
\label{Eq:OEPx_familiar}
  v_x(\vr)=-\sum_i^{occ}\! \int \! d\vrp \! \int \! d\vrpp \left[
            \phi_i^*(\vrp)G_i(\vrp,\vrpp)
	    \frac{\delta E_x}{\delta\phi_i^*(\vrpp)}+c.c.\right]
	    \chi_0^{-1}(\vrpp,\vrp)
\end{equation}
with
\begin{equation}
\label{Eq:def:G_i}
  G_i(\vrp,\vrpp)=\sum_{j\ne i}\frac{\phi_j(\vr)\phi_j^*(\vrp)}
                               {\epsilon_j-\epsilon_i} \quad.
\end{equation}
In OEPx calculations \emph{local} correlation is frequently added by 
including the LDA correlation energy 
\begin{equation}
\label{Eq:LDAcorr}
  E_c^{\rm LDA}[n]=\int \! d\vr \: \epsilon_c^{\rm HEG}(n(\vr))
\end{equation}
in the expression of the total energy  \refeq{Eq:KS_Etot_OEPx}.  
Here we follow the parametrization of Perdew and Zunger
\cite{Perdew/Zunger:1981} for 
the correlation energy density $\epsilon_c^{\rm HEG}(n(\vr))$ of the 
homogeneous electron gas (HEG) based on the data of Ceperley and Alder 
\cite{Ceperley/Alder:1980}.

Adding LDA correlation improves the ionisation potential of the constituent
atoms as will be shown later, but has little affect on the quasiparticle
bandstructures of the compound semiconductors presented in 
section \ref{Sec:26}.
In the following we
will refer to the OEP exact-exchange only scheme by OEPx in order to
distinguish it from the scheme with added LDA correlation, termed OEPx(cLDA).

The improvement of the exact-exchange approximation over the conventional LDA or
GGA approach is largely due to the removal of the self-interaction in the 
OEPx formalism. 
Or in other words, the interaction of an electron with
itself, as introduced by the Hartee potential \refeq{Eq:def:vHartree}, is fully
removed by the exact $v_x$. Since the OEPx exchange potential derives from an orbital
dependent functional it exhibits an integer discontinuity with respect to
variations in the particle number \cite{Krieger/Li/Iafrate:1992} (as does the
full OEP potential \cite{Casida:1999}), unlike the
smoothly varying exchange-correlation potentials in LDA or GGA.
For solids, however, it is still unclear how to interpret this discontinuity in 
OEPx bandstructure calculations \cite{Magyar/Fleszar/Gross:2004}. 
This shall be of no immediate concern for our work because using 
the OEPx eigenvalues as input for the Green's function is formally well defined.

On a more important note we like to emphasise that
the exact-exchange potential can be derived from
the framework of many-body perturbation theory unlike $v_{xc}$ in the 
LDA or GGA, for which
this is only true in the jellium limit.  
Equation \refeq{Eq:OEPx} therefore provides a rigorous connection between the 
exact-exchange Kohn-Sham system and $GW$. From this observation we draw the 
hypothesis that the OEPx approach provides a better starting point for 
quasiparticle energy calculations than LDA or GGA. 
In the remainder of this paper we will verify this hypothesis
numerically for the four materials considered here.

As a side note we like to mention that part of the quasiparticle screening
can already be incorporated on the level of DFT. Rooted 
in the  generalised Kohn-Sham scheme \cite{GDFT:1993},
the so called screened-exchange 
approximation (sX-LDA) 
\cite{GDFT:1993,Kleinman/Bylander:1990,Lento/Nieminen:2003,
                    Kim/Zhao/Freeman/Mannstadt:2004} 
utilises an \emph{ad hoc} model (typically Thomas-Fermi) to screen the
exchange-interaction. Conceptionally the sX-LDA
approach is similar to the screened-exchange (SEX) approximation in
the framework of the $GW$ method \cite{Hybertsen/Louie:1986}. 
Bandstructures in the
sX-LDA scheme are reported to be in good agreement with experiment for 
semiconductors  and insulators \cite{GDFT:1993,Kleinman/Bylander:1990,
                 Kim/Zhao/Freeman/Mannstadt:2004}. This indicates that sX-LDA 
could be a viable alternative to the OEPx as a starting point for $GW$
calculations. A numerical verification of this conjecture, however,
will be left to future studies. 

\section{Shallow Semicore $d$-Electron Systems}
\label{sec:d-electrons}

The II-VI compounds and most of the group-III-nitrides are characterised by 
the semicore $d$-electrons introduced by the cation.
Compared to other materials with semicore states the cation $d$-electrons in 
ZnO, ZnS and CdS  have a small binding energy of around 9 eV 
\cite{Weidemann/Becker:1992,Ley/Shirley:1974} and are thus
energetically close to the valence states with mostly anion
$p$-character. For GaN X-ray photoemission experiments report two peaks at
17.7 eV and 14.2 eV that have been attributed to the Ga 3$d$ states and
the N $2s$ bands, respectively \cite{Ding/etal:1996}
(see section \ref{Sec:BS} for bandstructure plots).
In the Zn and Cd based semiconductors the $p$-$d$ hybridisation is therefore
larger than in GaN, which in turn exhibits stronger $s$-$d$ coupling. 
The effect of 
this qualitative difference on the electronic wavefunctions and densities
will be discuss in section \ref{Sec:dens_wfc}.

In light of this argument it is essential to find a good description
of the $p$-$d$ and $s$-$d$ coupling in order to calculate the structural and
the electronic properties of these materials. 
In the past a common approximation has been to 
simply remove the $d$-electrons 
from the calculation and to model the interaction of the valence electrons with
the atomic core by a pseudopotential (see next section). Although this is
computationally very efficient it leads to a distinct disagreement between
theory and experiment for the strucural properties
\cite{Fiorentini/Methfessel/Scheffler:1993,Schroer/Krueger/Pollmann:1992,
      Schroer/Krueger/Pollmann:1993} 
as well as for electronic excitations
\cite{Rohlfing/Pollmann:1995,Rohlfing/Pollmann:1998,Luo/Louie:2002}.
 
In the following section we will introduce the pseudopotential approximation 
and elucidate how shallow semicore electron systems
can be treated consistently in our OEPx(cLDA)+$GW$ approach for the 
electronic structure.

\subsection{Core-Valence Partitioning and Pseudopotentials}
\label{Sec:PPs}

The \emph{pseudopotential} approach to electronic structure methods for
polyatomic systems employs the more general 
concept of \emph{core-valence partitioning}. Motivated by the observation that
deep core states are relatively inert and do not contribute to chemical bonding
they are often treated on a different footing than valence or 
semicore\footnote{In the destinction between valence and core states,
\emph{semicore} states have
binding energies between those of core and valence electrons, but
hybridize with valence states or contribute to chemical bonding.}
electrons. 

In \emph{ab initio} pseudopotential calculations the potential due to
the nuclei and the core electrons is replaced by an 
\emph{ionic pseudopotential}. A good pseudopotential should be smooth and 
should describe
the interaction with the remaining valence electrons of the atom well, 
while at the same time being transferable across chemically different 
environments. The pseudopotential is constructed by replacing
the valence wavefunctions of the isolated atom by smooth, nodeless 
pseudowavefunctions inside a given cutoff radius. Inversion of the atomic
Schr\"odinger equation then yields the atomic pseudopotential and by
substracting out the Hartree and exchange-correlation potential generated by the
pseudovalence states the ionic pseudopotential is obtained.
For reciprocal-space electronic structure methods a substantial 
reduction of the computational cost can be gained in this way, because the 
oscillations of the atomic wavefunctions near the nucleus no longer have to be 
resolved  by plane-waves. Once the pseudopotential has been generated the core
electrons have been removed from the calculation of the polyatomic system
and remain frozen inside the nucleus. Core relaxation and polarisation
effects can thus not be taken into account, because the pseudopotential is 
typically not adjusted  to the new chemical environment during the 
self-consistency cycle.

Since any calculation following core-valence 
partitioning can never be better than the accuracy with which the interactions 
between core and valence electrons have been 
treated \cite{Shirley/Martin/Bachelet/Ceperley:1990}, it has long been 
recognised that consistency is paramount.
For DFT calculations with local or semi-local exchange-correlation functionals
this is achieved by employing the same functional in the generation of
the pseudopotential and the calculation of the extended or molecular system. In
fact \emph{ab initio} LDA or GGA pseudopotentials are now routinely 
applied in LDA or GGA calculations to a wide range of systems. 
When going beyond DFT, however, consistency will almost inevitably  be 
violated. In quantum Monte Carlo calculations for example Hartree-Fock
pseudopotentials are frequently employed \cite{Lee/Needs:2003}, whereas $GW$
calculations are almost exclusively based on an LDA or GGA ground state 
\cite{Aryasetiawan/Gunnarsson:1998}
and the respective pseudopotential \cite{Aulbur/Jonsson/Wilkins:2000}. 
One way of compensating for the
lack of many-body pseudopotentials would be to introduce core polarisation effects
into the pseudopotential \cite{Lee/Needs:2003,Shirley/Martin:1993}. 
By extending the $GW$ formalism to include core 
contributions in the dielectric screening and the self-energy such
core-polarisation-potentials (CPP) have also been incorporated successfully
into the $GW$ method \cite{Shirley/Zhu/Louie:1997}. 

Ultimate consistency would of course imply to abolish 
pseudopotential-core-valence partitioning and to treat all electrons on the 
same footing. While DFT-all-electron methods have become a standard technique
in condensed matter physics all-electron $GW$ implementations 
are only slowly emerging. Early calculations were carried out in the atomic
sphere approximation (ASA) to the linearised muffin-tin-orbital (LMTO) method
\cite{Aryasetiawan/Gunnarsson:1995,Aryasetiawan/Gunnarsson:1996,
      Oshikiri/Aryasetiawan:1999,Oshikiri/Aryasetiawan:2000}.
Only fairly recently $GW$ calculations in the 
full-potential (FP-LMTO)
\cite{Kotani/Schilfgaarde:2002,Faleev/Schilfgaarde/Kotani:2004} or the
full-potential linearised augmented-plane-wave (LAPW) method
\cite{Usuda/Schilfgaarde:2002,Usuda/Hamada/etal:2004,
      Ku/Eguiluz:2002} have been reported, 
whereas the projector-augmented-wave (PAW) scheme
\cite{Arnaud/Alouani:2000,Lebegue/Arnaud/Alouani/Bloechl:2003}  falls into a mixed
category. While the effect of the core-electrons on the valence electrons is 
included in the PAW method, the augmentation projectors are not updated in the 
calculation.
That puts the treatment of the core electrons in PAW on the same footing as in 
a pseudopotential approach.

In order to reduce the size of the frozen core in the pseudopotential 
approach and therefore to move towards an all-electron description,  
 more core electrons can be explicitly considered 
as valence electrons in the calculation
\cite{Rohlfing/Pollmann:1995,Rohlfing/Pollmann:1998,
      Luo/Louie:2002,Tiago/Ismail-Beigi/Louie:2003,Kralik/Chang/Louie:1998}. 
Special care has to be taken, however, if an angular momentum channel has 
more than one bound state. 
The computational costs for treating core states in this fashion are 
moderate when localised basis sets are used 
\cite{Rohlfing/Pollmann:1995,Rohlfing/Pollmann:1998},
but fromidable in a plane-wave implementation 
\cite{Luo/Louie:2002,Tiago/Ismail-Beigi/Louie:2003,Kralik/Chang/Louie:1998}.

\begin{figure}[ht]
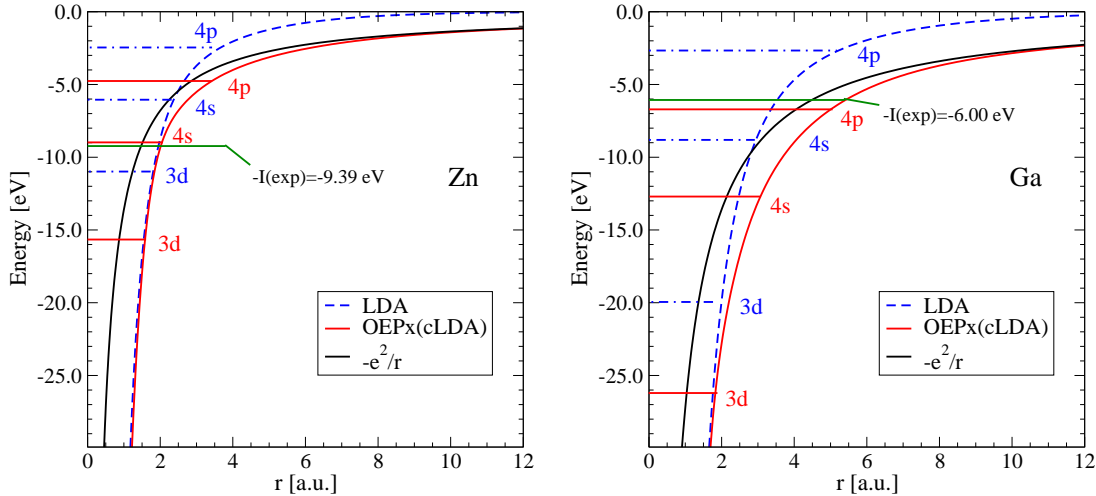

  \begin{center}
    \begin{minipage}[ht]{0.48\linewidth}
      \epsfig{bbllx=14,bblly=43,bburx=517,bbury=514,clip=,
	      file=graphs/Zn-3d_EXX_atom_eV.eps,width=1.0\linewidth}
    \end{minipage}
    \quad
    \begin{minipage}[ht]{0.48\linewidth}
      \epsfig{bbllx=14,bblly=43,bburx=517,bbury=514,clip=,
	      file=graphs/Ga-3d_EXX_atom_eV.eps,width=1.0\linewidth}
    \end{minipage}
  \end{center}

    \caption{\label{fig:Zn_Ga_atom}\small{
             Effective Kohn-Sham potential for the neutral Zn (left panel) and 
	     Ga (right panel) atom: 
	     the OEPx(cLDA) potential (red line) reproduces the correct
	     asymptotic decay $-e^2/r$ (black line), whereas the LDA (blue,
	     dashed line) decays exponentially and thus underbinds the electrons.
	     The atomic levels (shown as horizontal lines) are lowered in the
	     OEPx(cLDA) approach compared to the LDA resulting in good
	     agreement with the experimentally measured ionisation potential
	     (green horizontal line).}}
\end{figure}

Since the quality of the basis set in real and reciprocal space methods 
can be systematically monitored and 
increased simply by including more points or plane-waves we prefer to
stay within the framework of
plane-wave and so called \emph{mixed-space} electronic structure calculations
\cite{Rojas/Godby/Needs:1995,GW_space-time_method:1998} 
in this study.
In order to capitalise on the ``cost-effectiveness'' of the pseudopotential
approach and the numerical convenience of plane-waves we combine the $GW$
excited state calculations with DFT ground state calculations
in the exact-exchange formalism (denoted OEPx+$GW$ in order to 
draw a distinction to $GW$ calculations based on the LDA ground state,
LDA+$GW$, which we perform for comparison). Employing
exact-exchange pseudopotentials \cite{Moukara/etal:2000,Engel/etal:2001}
allows us to freeze the $s$ and $p$ electrons of the semicore $d$-shell in the
core of the pseudopotential while retaining the $d$-electrons as valence 
states in the calculation (see also Tab. \ref{Tab:PPPar}). 
Our numerical results for the quasiparticle bandstructures in the OEPx+$GW$
formalism (sections \ref{Sec:BS}, \ref{Sec:26} and \ref{Sec:g3}) 
show good agreement with available photo-electron spectroscopy data,
which indicates that the dominant interaction between the core and valence
electrons is exchange mediated and is well described by the exact-exchange 
pseudopotentials for the compounds considered here. 
This is one of the key results of our work and we will elucidate it further in
later sections.

\begin{table}
  \caption{\label{Tab:PPPar}{\small 
           Pseudopotential reference parameters: The electron configuration 
	   is given in the 2nd column with the core electron configuration in 
	   square brackets. The same core radius $r_c$ (given in bohr) 
	   was used for all angular
	   momentum components and the local component $l_{loc}$ is listed in
	   the 4th column. For $N$ only $s$ and $p$ components are considered 
	   \cite {Stampfl/VandeWalle:1999}.
	   }}
  \begin{indented}
  \item[]
    \begin{tabular}{llcc|llcc}
       \br
       \multicolumn{1}{c}{Cation} &
       \multicolumn{1}{c}{Configuration} &
       \multicolumn{1}{c}{$r_c$ } &
       \multicolumn{1}{c|}{$l_{loc}$} &
       \multicolumn{1}{c}{Anion} &
       \multicolumn{1}{c}{Configuration} &
       \multicolumn{1}{c}{$r_c$} &
       \multicolumn{1}{c}{$l_{loc}$} \\ \br
          Zn &  [Ar]3$d^{10}$4$s^2$       & 2.2 & $s$ &
          N  &  [He]2$s^2$2$p^3$          & 1.5 & $p$ \\	  
          Ga &  [Ar]3$d^{10}$4$s^2$4$p^1$ & 2.2 & $s$ &
          O  &  [He]2$s^2$2$p^4$          & 1.6 & $d$ \\	  
          Cd &  [Kr]4$d^{10}$5$s^2$       & 2.2 & $s$ &
          S  &  [Ne]3$s^2$3$p^4$          & 1.7 & $d$ \\	  
	   \br
    \end{tabular}
  \end{indented}
\end{table}

Figure \ref{fig:Zn_Ga_atom} illustrates the improvement
for the Kohn-Sham eigen-energies obtained in the OEPx(cLDA) formalism
compared to the conventional LDA treatment 
for the Zn (on the left) and the Ga atom (on the right). 
In our work we follow the approach of 
Mourkara \emph{et al}. \cite{Moukara/etal:2000} in constructing the OEPx and 
OEPx(cLDA) pseudopotentials. 
The parameters for the LDA and exact-exchange pseudopotentials are 
taken to be the same and are listen in Table \ref{Tab:PPPar}.
Returning to Fig. \ref{fig:Zn_Ga_atom} we observe that
the effective Kohn-Sham potential (red line) follows the correct 
asymptotic $-e^2/r$ potential
outside the atom (black line), whereas the the LDA potential (blue, dashed line) 
decays exponentially fast. The strong underbinding of the electrons inherent to
the LDA is greatly reduced in the OEPx(cLDA) approach. This is largely due to 
the removal of the
self-interaction in the OEPx(cLDA), which leads to a lowering of the atomic 
levels. Good agreement between the highest occupied electronic level and the
experimental ionisation potential is then observed.

We will close this section with a brief overview of the compuational cost
associated with our approach. For this purpose we will focus on the 
example of zinc-blende GaN. To converge the LDA full-shell pseudopotential 
ground state calculation, where the Ga 3$s$ and 3$p$ are treated as valence 
electrons, a plane wave cutoff of 300 Ry is needed. To achieve
convergence of better than 0.1 eV in the quasiparticle energies a cutoff of
144 Ry for the correlation and 270 Ry for the exchange part of the $GW$ self-energy
is required
\cite{GaN_unp}\footnote{This compares well to parameters for
other materials \cite{Tiago/Ismail-Beigi/Louie:2003,Kralik/Chang/Louie:1998}, 
although the parameters in Ref. \cite{Luo/Louie:2002} appear to be somewhat low.}.
For the computationally more intensive correlation part this is twice as much 
as is needed when the Ga 3$s$ and 3$p$ electrons are frozen in the core of the 
pseudopotential (see Tab. \ref{Tab:CompPar}).
Since the $GW$ space-time method scales quadratically with the number of
real-space points and cubically with the number of $\vG$-vectors doubling the
cutoff increases the computational load by a factor that lies somewhere between
8 and 22 depending on how much the inversion of the dielectric function
dominates the scaling. It is
therefore desirable to keep the cutoff in plane wave $GW$ calculations as low as
possible.

A $GW$ plane wave cutoff of 70 Ry for GaN can still be considered as moderate
both in terms of memory and hard disk space usage as well as computational time. 
The OEPx+$GW$ calculations for the zinc-blende structures presented here
can still be carried out on a modern PC or workstation, whereas the 
memory and disk space requirements for the full-shell $GW$ calculations quickly
reach dimensions presently only available on high-performance computing 
clusters and at super computer facilities. 

The computationally most intensive part of our
OEPx+$GW$ approach are currently the exact-exchange calculations. Since we
follow the OEP scheme for solids developed by G\"orling \cite{Goerling:1996} an
explicit calculation and inversion of the static polarisability is also required
for the OEPx ground state calculations (see equation \refeq{Eq:OEPx_familiar}). 
In our experience every iteration of the
OEPx self-consistency cycle in this implementation 
is comparable to a full $GW$ calculation in terms 
of hardware requirements and run time. Altogether between 5 and 8 OEP cycles are
needed to converge the exact-exchange calculations presented in this article.
Recently, however, a new OEP scheme has been proposed
\cite{Kuemmel/Perdew:2003a,Kuemmel/Perdew:2003b,Kuemmel/Kronik/Perdew:2004},
which circumvents an explicit solution of equation \refeq{Eq:OEPx_familiar}. 
This alternative scheme promises to be
computationally much more efficient than our current implementation and
therefore to substantially reduce the cost of the OEPx+$GW$ approach.

\subsection{Computational Details}
All ground state calculations are performed with the pseudopotential
plane-wave DFT 
code \texttt{SFHIngX} \cite{SFHIngX}. The OEPx formalism for solids 
\cite{Goerling:1996} has recently been added \cite{EXXpaper:2004} to this 
program package. 
For the DFT calculations in the 
local-density approximation \cite{Kohn/Sham:1965} we use 
the parameterisation by Perdew and Zunger \cite{Perdew/Zunger:1981} of the
Ceperley and Alder data for the homogenous electron gas 
\cite{Ceperley/Alder:1980}. 

All LDA pseudopotentials are constructed with the
\texttt{fhi98PP} generator \cite{fhi98PP}, in the Troullier-Martins (TM)
scheme \cite{Troullier/Martins:1991} and transformed to the non-local, separable
Kleinman-Bylander form \cite{Kleinman/Bylander:1982}. 
For the four compounds presented in this article our
LDA calculations with these pseudopotentials reproduce
the  bandstructure of all-electron LDA  calculations in the 
(linearised) augmented-plane-wave plus local orbital ((L)APW+lo) approach
\cite{Wien2k} to within 0.1 eV \cite{Ricardo}.
For the OEPx and OEPx(cLDA) pseudopotentials we follow the method developed
by Moukara \emph{et al}. \cite{Moukara/etal:2000}.

\begin{table}
  \caption{\label{Tab:CompPar}{\small 
           Computational parameters: zinc-blende lattice constant ($a_{ZB}$)
	   in \AA, plane-wave cutoff ($E_{cut}$), reduced
	   cutoff for the inversion of $\chi_0$ in OEPx
	   ($\chi_{cut}$) and the band cutoff for the Green function
	   ($b_{cut}$) are listed in Rydberg. Column 5 gives
	   the change in the macroscopic dielectric constant 
	   ($\varepsilon_M$) in percent
	   when the non-local (NL) component of the OEPx(cLDA) pseudopotential is
	   included and the last column the resulting change in the 
	   quasiparticle band gap.}}

  \begin{indented}
  \item[]
    \begin{tabular}{crrrrrr}
       \br
       \multicolumn{1}{c}{} &
       \multicolumn{1}{c}{$a_{ZB}$} &
       \multicolumn{1}{c}{$E_{cut}$} &
       \multicolumn{1}{c}{$\chi_{cut}$} &
       \multicolumn{1}{c}{$b_{cut}$} &
       \multicolumn{1}{c}{$\Delta\varepsilon_M/\varepsilon_M$} &
       \multicolumn{1}{c}{$E_{gap}^{NL}-E_{gap}$} \\ \br
          GaN & 4.500 & 70 Ry & 45 Ry & 40 Ry &  -10 \%&  0.07 eV\\ 
	  ZnO & 4.620 & 60 Ry & 35 Ry & 56 Ry &  -15 \%&  0.15 eV\\
	  ZnS & 5.400 & 60 Ry & 35 Ry & 40 Ry &    0 \%&  0.00 eV\\
	  CdS & 5.818 & 50 Ry & 30 Ry & 24 Ry &   +8 \%& -0.03 eV\\ \br
    \end{tabular}
  \end{indented}
\end{table}

The $GW$ calculations are performed employing the $GW$ space-time approach
\cite{Rojas/Godby/Needs:1995} in the \texttt{gwst} implementation 
\cite{GW_space-time_method:1998,GW_space-time_method_enh:2000}. 
To accelerate
the convergence with respect to the number of k-points and to avoid numerical
instabilities arising from the Coulomb singularity at $\vk=0$ in reciprocal
space we treat head and wings of the dielectric matrix \refeq{Eq:eps_RPA} 
in  $kp$-perturbation theory 
\cite{Hybertsen/Louie:1986,Baroni/Resta:1986,GW_space-time_method:1998}. 
In this approach the inverse of
the head of the inverse dielectric matrix is then equivalent to the macrosopic
dielectric constant $\varepsilon_M$ in the RPA.
We find that $\varepsilon_M$ is fully converged if
a 4$\times$4$\times$4 $\vk$-grid with an offset of 
$[\frac{1}{2},\frac{1}{2},\frac{1}{2}]$ is used for the Brillouin zone
integration.  A  4$\times$4$\times$4 $\vk$-grid with no offset then proves
to be sufficient for the full $GW$ calculation.
The non-local part of the OEPx(cLDA) pseudopotentials, which is
fully taken into account in our implementation, reduces the
RPA macroscopic dielectric constant $\varepsilon_M$ in GaN and ZnO, but increases
it in CdS, as can be seen in Tab. \ref{Tab:CompPar}. 
The quasiparticle band gap, however, is modified only slightly. 
More details will be given elsewhere \cite{GWtechnical}.

With respect to the plane-wave cutoff ($E_{cut}$) in all DFT and $GW$ calculations
the single particle energies are converged to better than
a tenth of an eV for the values listed in Tab. \ref{Tab:CompPar}.
The static polarisability in the OEP calculations 
is set up and inverted in a smaller plane-wave basis with a reduced 
cutoff energy $\chi_{cut}$. Further increasing $\chi_{cut}$ changes the
eigenvalues by less than 0.01 eV \cite{EXXpaper:2004}.
Finally all unoccupied bands with energy $\epsilon-E_F< b_{cut}$ are 
included in the $GW$ Green's function. All calculations are carried out for
zinc-blende structures. In order to facilitate an unambigious comparison with
experiment or in other words to benchmark the performance of our computational
approach against the exact theory all calculation are carried out at 
the experimental lattice constants, as 
reported in column 1 of Tab. \ref{Tab:CompPar}.

To solve the quasiparticle equation \refeq{Eq:qp_eq} we approximate the
quasiparticle wavefunctions by the Kohn-Sham eigenfunctions:
$\psi_{n\vk}(\vr)=\phi_{n\vk}(\vr)$. For the upper valence and conduction bands
of standard semiconductors numerical investigations indicate that this
approximation is well justified 
\cite{Aulbur/Jonsson/Wilkins:2000,Aryasetiawan/Gunnarsson:1998}, but it
breaks down for certain surface 
\cite{White/Godby/Rieger/Needs:1997,Rohlfing/Wang/Krueger/Pollmann:2003,
      Fratesi/Brivio/Rinke/Godby,Fratesi/Brivio/Molinari}
and cluster states
\cite{Pulci/Reining/Onida/DelSole/Bechstedt:2001,
      ClusterImStates:2004}.
We will leave an analysis of the quasiparticle wavefunctions to future studies
and instead solve the diagonal quasiparticle equation 
\begin{equation}
\label{Eq:qpe}
  \epsilon_{n\vk}^{qp}=\epsilon_{n\vk}^{\rm DFT}+\bracket{\phi_{n\vk}}
                       {\Sigma_{xc}^{GW}(\epsilon_{n\vk}^{qp})-v_{xc}-\Delta\mu}
		       {\phi_{n\vk}}
\end{equation}
iteratively for the quasiparticle energies $\epsilon_{n\vk}^{qp}$.
At every iteration step the DFT energies are shifted by a constant 
$\Delta\mu$ that aligns the Fermi energies before
and after applying the $GW$ self-energy corrections. This makes the
solution of the quasiparticle equation robust against different energy zeros
of the exchange-correlation potential, in particular if the energy zero is 
not the same as that of the self-energy. The alignment constant $\Delta\mu$ was 
first introduced by
Hedin \cite{Hedin:1965} for the electron gas to simulate to some extent the 
effects of self-consistency in $G$. Later it was shown for a model system
\cite{Schindlmayr:1997} that $\Delta\mu$ is instrumental in keeping 
violations of charge conservation negligible.

\subsection{Electronic Structure of II-VI Compounds and Group-III-Nitrides}

\subsubsection{Electron Density and Wavefunctions}
\label{Sec:dens_wfc}

\begin{figure}
  \begin{center}
    \includegraphics[
          width=0.98\columnwidth]{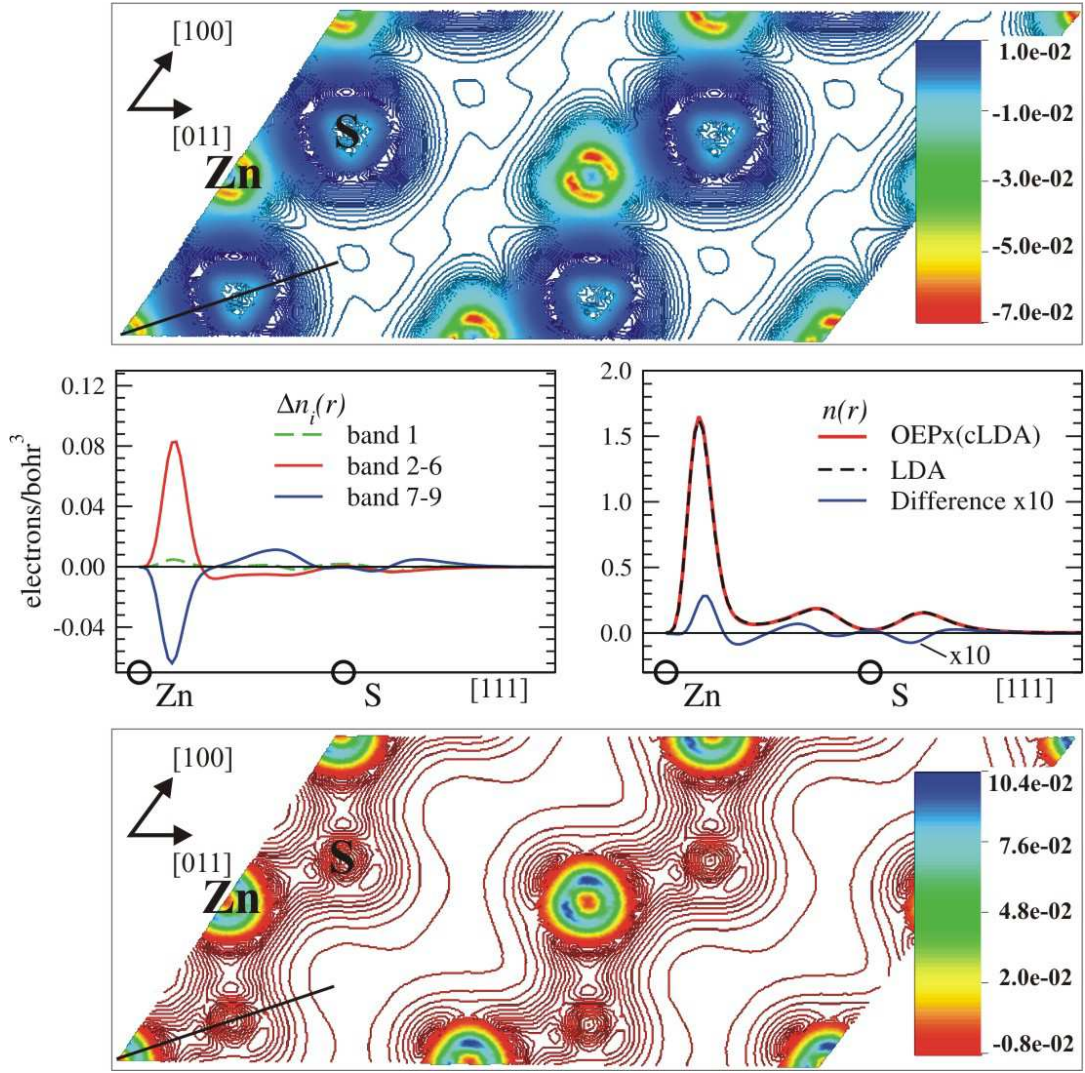}
    \caption{\label{fig:ZnS-density-diff} 
             Comparison between OEPx(cLDA) and LDA results for the
	     electron density and the partial densities difference
	     ($\Delta n_i(\vr)$) of ZnS:
	     Positive density differences indicate
	     an accumulation in OEPx(cLDA). 
	     \emph{Middle panel -- left}: 
	     partial density differences 
	     along the [111] direction through the unit cell;
	     while the $s$-band (green, dashed) remains largely unaffected,
	     the $d$-electrons (red) are drawn from the valence region and are
	     localised more strongly on the Zn atom, whereas the remaining
	     valence electrons (blue) accumulate stronger in the bonding region
	     in OEPx(cLDA).
	     \emph{Middle panel -- right}: 
	     The OEPx(cLDA) (red, solid) and LDA (black, dashed) 
	     electron densities are 
	     almost indistinguishable. The
	     density difference (blue line -- magnified by a 
	     factor of 10) reflects the partial density changes. 
	     (The scale of the ordinate is the same as in 
	     Fig. \ref{fig:GaN-density-diff}.)
	     Cross sections through the partial density difference in
	     electrons/bohr$^3$ for valence 
	     bands (7-9) and $d$-bands (2-6) are displayed in the \emph{top} 
	     and \emph{bottom panel}, respectively. The path taken for the
	     one-dimensional plots in the middle panels is marked by the 
	     black line.
	     }
  \end{center}
\end{figure}

\begin{figure}
  \begin{center}
    \epsfig{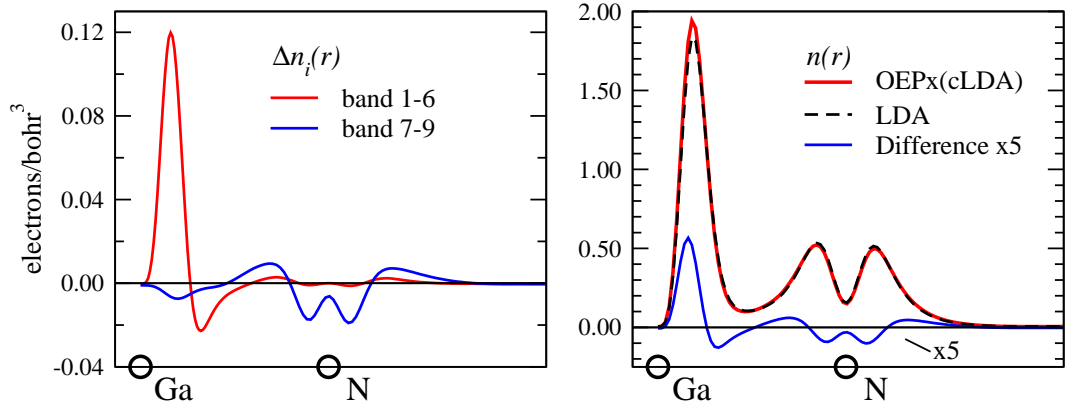}
    \caption{\label{fig:GaN-density-diff} 
             \emph{Left hand side}: 
	     partial density differences 
	     ($\Delta n_i(\vr)$)
	     along the [111] direction through the unit cell of GaN
	     (same as Fig. \ref{fig:ZnS-density-diff}).
	     The $d$-electrons (red) localise stronger on the Ga atom, but are
	     not drawn from the remaining 
	     valence electrons (blue) as in ZnS.
             \emph{Right hand side}: 
	     The OEPx(cLDA) (grey) and LDA (black) electron densities are 
	     very similar. The
	     density difference (red line -- magnified by a 
	     factor of 5) reflects the stronger localisation of the
	     $d$-electrons. 
	     }
  \end{center}
\end{figure}

Approximating the quasiparticle wavefunctions with the Kohn-Sham 
eigenfunctions introduces a dependence on the exchange-correlation functional 
into the $GW$ calculation.
Changing the functional from
LDA to OEPx(cLDA) in the ground state calculation 
will modify the wavefunctions and
subsequently alter the interaction of the quasiparticles. 
To elucidate these changes we have
plotted the charge densities and charge density difference for 
ZnS and GaN along the [111] direction through the unit cell
(middle panel on the right of \mbox{Fig. \ref{fig:ZnS-density-diff}} 
and \ref{fig:GaN-density-diff}, respectively). In the panel on the left
hand side the differences are broken down in terms of the partial charge 
densities $\Delta n_i(\vr)$ that have
been obtained by summing over all wavefunctions in the bands indicated.
In Ga the 3$d$ electrons are lower in energy than in Zn  
(see Fig. \ref{fig:Zn_Ga_atom}). The $d$-band complex is therefore found
closer to the N 2$s$ states in GaN than to the S 3$s$ in ZnS, as the
bandstructures in Fig. \ref{fig:ZnS_BS_b} and \ref{fig:GaN_BS} illustrate.
For ZnS we can thus clearly distinguish between 
the partial density of bands with
mostly S 3$s$ (dashed green line), Zn 3$d$ (red line) and $sp$ character (blue
line), whereas for GaN we included the N 2$s$ states in the sum over the 
$d$-bands (red line). To visualise the
spatial variation of the partial electron density difference we included 
cross-section plots for the $sp$-valence states (top panel) and
$d$-states (lower panel) of ZnS in Fig. \ref{fig:ZnS-density-diff}.

Inspection of \mbox{Fig. \ref{fig:ZnS-density-diff}}
and \ref{fig:GaN-density-diff} shows that
the electron densities in LDA and OEPx(cLDA) are nearly identical and almost
indistinguishable on the scale of the plot. A magnification of the density
difference, however, reveals small charge accumulations in the bonding
region and on the cation for the OEPx(cLDA) ground state compared to the 
LDA one. Comparing GaN with ZnS the charge localisation on the Ga atoms 
in the OEPx(cLDA) approach is found to be more pronounced than on the Zn atoms.
A closer look at the partial densities elucidates that 
this qualitative difference arises from the
different $d$-electron hybridisation in these two compounds. 
In ZnS the $d$-states hybridise with the
$sp$-valence states so that the removal of self-interaction in the OEPx(cLDA)
leads to a stronger
localisation of the $d$-electrons (red line) on the Zn atom, whereas the 
valence electrons (blue line)
are drawn from Zn into the bonding region. The two effects are of opposite
nature and as a result a negligible overall charge 
density difference is observed since
the S 3$s$ states remain almost unaffected by the change in the
exchange-correlation potential. This behaviour is analogously found in 
ZnO and CdS.

In GaN on the other hand the $d$-states hybridise
strongly with the N 2$s$ states (shown together as red line) 
and are energetically separated from
the remaining valence electrons (blue line). Again we observe a localisation of 
the $d$-electrons on the cation upon removal of the self-interaction. But this
time the charge accumulation is not compensated by a reduction of the valence 
electrons, which were
already bound closer to the N atom and are now shifted more into the bonding 
region by the OEPx(cLDA).

Previously it was observed that
LDA densities are more homogeneous than their OEPx counterpart 
\cite{Staedele/etal:1999}. Furthermore, 
the OEPx density was found to localise stronger in the bonding region for 
GaAs and GaN \cite{Aulbur/Staedele/Goerling:2000}. In these calculations 
the $d$-electrons were frozen in the core of the pseudopotential. 
Although these observations certainly apply to the \emph{upper valence} 
electrons our results
show that the $d$-electrons introduce the opposite effect. This leads 
to a more pronounced electron localisation on the cation in GaN in the 
OEPx approach\footnote{The inclusion of LDA correlation in OEPx(cLDA) does not 
change
these observations.} and cancels out the charge accumulation in the bonding 
region in ZnO, ZnS and CdS. 

In the following sections we will analyse the implications of these
observations for the bandstructure of the four 
semiconductors and compare to spectroscopical data where available.

\subsubsection{Band Gaps}    
\label{Sec:BS}
\begin{figure}
  \begin{center}
    \includegraphics[width=0.5\columnwidth]{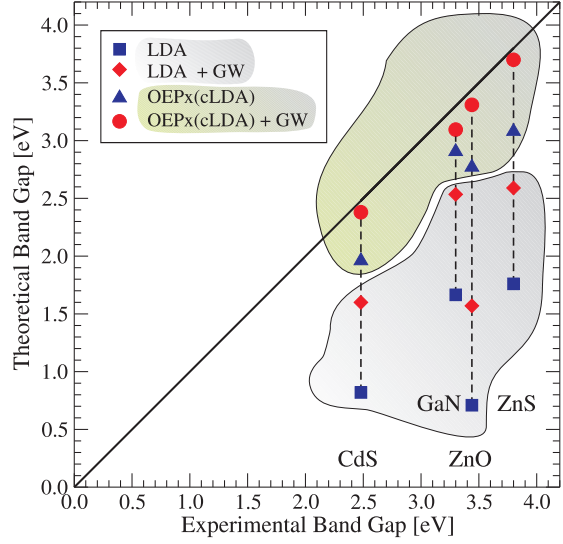}
    \caption{\label{fig:bandgaps} 
             Theoretical versus experimental band gaps: the OEPx(cLDA) based
	     schemes systematically open the band gap compared to the LDA 
	     based calculations. Our OEPx(cLDA)+$GW$ calculations with the
	     cation $d$-electrons included as valence electrons agree very well
	     with the experimental values 
	     (see Tab. \ref{tab:qpe_bg} for references). (For ZnO an estimate of 0.2 eV was added to the
	     zinc-blende values in order to compare to the experimental results
	     for wurtzite (see text).) }
  \end{center}
\end{figure}

Before we proceed with
a more detailed analysis of the electronic structure for the selected
II-VI compounds and group-III-nitrides in our approach
we like to highlight one of our key results: 
\emph{the OEPx(cLDA) based schemes systematically open the band gap compared to the
LDA based variants}, as illustrated in Fig. \ref{fig:bandgaps}.
\emph{Our $GW$ bandstructure calculations
reproduce the experimental values very 
well}\footnote{For ZnO experimental data is only available 
for the wurtzite structure. On the level of LDA the band gap is 0.2 eV larger in
wurtzite than in zinc-blende. To compare with the experimental data
we have therefore adjusted all values for ZnO in Fig. \ref{fig:bandgaps} by
this amount.} \emph{when starting from the OEPx(cLDA) ground state.}
In the LDA based $GW$ calculations on the other hand the band gaps are 
underestimated appreciably with the LDA itself giving the most
severe underestimation
\cite{Fiorentini/Methfessel/Scheffler:1993,Schroer/Krueger/Pollmann:1992,
      Schroer/Krueger/Pollmann:1993}. 
      
\subsubsection{II-VI Semiconductors}
\label{Sec:26}
\begin{figure}[ht]
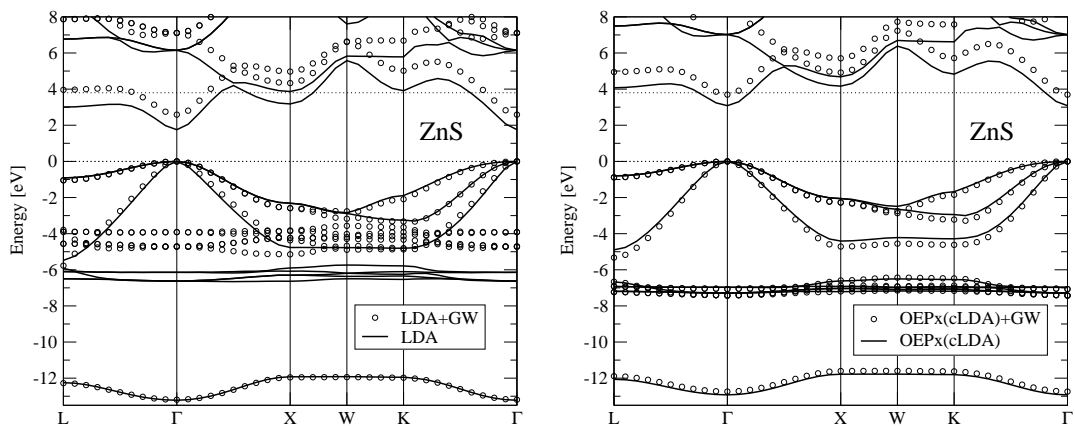

  \begin{center}
    \begin{minipage}[ht]{0.47\linewidth}
      \epsfig{bbllx=39,bblly=71,bburx=512,bbury=453,clip=,
	      file=graphs/ZnS_BS_LDA_GW.eps,width=1.0\linewidth}
    \end{minipage}
    \quad
    \begin{minipage}[ht]{0.47\linewidth}
      \epsfig{bbllx=39,bblly=71,bburx=512,bbury=453,clip=,
              file=graphs/ZnS_BS_EXX_GW.eps,width=1.0\linewidth}
    \end{minipage}
  \end{center}

  \caption{\label{fig:ZnS_BS_b}\small{
           Band structure of ZnS in
	   LDA and LDA+$GW$ (panel on the left) compared with
	   OEPx(cLDA) and OEPx(cLDA)+$GW$ (panel on the right). 
	   Consistent pseudopotentials are used.
	   All bandstructures have been aligned at the valence band maximum
	   (dotted line at 0 eV). For reference the experimental band gap is
	   marked by the 2nd dotted line.}}
\end{figure}

In Fig. \ref{fig:ZnS_BS_b} we compare the four different computational 
approaches for
the whole bandstructure of ZnS\footnote{The results for
ZnS are representative for ZnO and CdS and we therefore only show 
bandstructure for ZnS.}.
To facilitate a numerical comparison we have also listed the 
band gaps and $d$-electron binding energies
in Tab. \ref{tab:qpe_bg} and \ref{tab:dbands} together with the values for the
other compounds. 

We observe that the reduction of the $p$-$d$ hybridisation due to the removal 
of the self-interaction in OEPx(cLDA) discussed in the
previous section  decreases the valence bandwith and a larger gap between 
the $p$ and $d$ bands is opened (right panel). 
The $d$-bands, however, remain almost in the same position as in the LDA.
The $GW$ valence bandstructure is very similar to the OEPx(cLDA) one,
apart from small dispersive shifts in the $p$-band complex. 
The conduction bands
are shifted up almost uniformly opening the fundamental band gap to 3.7 eV
compared to the experimental value of 3.8 eV.

Starting from the LDA ground state,
however, we find that in $GW$ the $d$-bands overlap in
energy with the $p$-bands  (left panel). 
The reason for this unphysical behaviour can be traced back to
the pseudopotential approximation.
Since the atomic $d$-orbitals of Zn and Cd
overlap considerably with the wavefunctions of the $s$ and $p$ electron in the 
same shell, the exchange interaction between the $d$ and the
remaining core electrons in this shell is strong. 
The poor performance
of the LDA can thus largely be explained by 
spurious self-interaction effects, as alluded to before and demonstrated
numerically by applying 
self-interaction corrections (SIC)
\cite{Vogel/Krueger/Pollmann:1995,Vogel/Krueger/Pollmann:1996,
      Vogel/Krueger/Pollmann:1997,Qteish:2000}.

\begin{figure}[ht]
  \begin{center}
    \epsfig{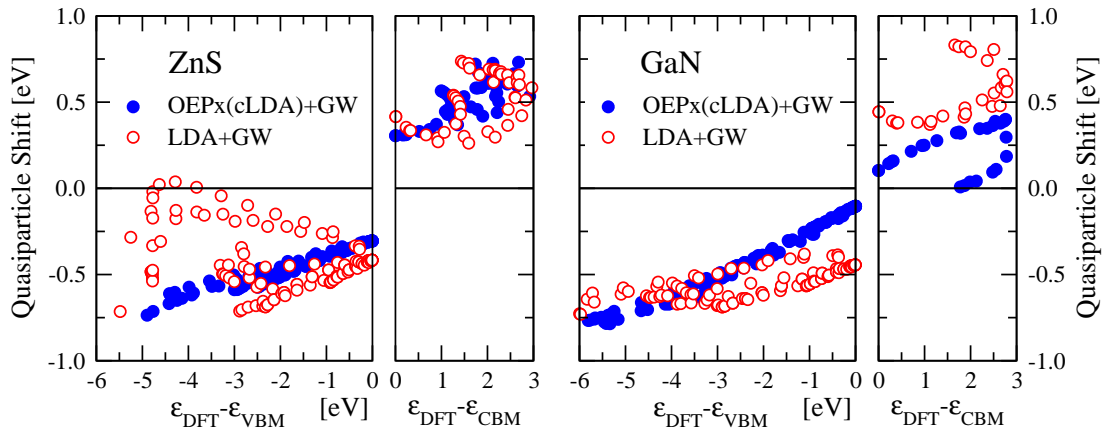}
  \end{center}
  \caption{\label{fig:ZnS_GaN_qpeshifts}\small{
           Quasiparticle shifts as function of the Kohn-Sham energy for ZnS and
	   GaN for $GW$ calculations based on an LDA (open circles) and an
	   OEPx(cLDA) (filled circles) ground state. In order to avoid
	   visual distortions due to the band gap difference between 
	   LDA and OEPx(cLDA) the curves have been aligned on the abscissa 
	   such that both the valence band maximum (VBM)
	   and the conduction band minimum (CBM) are equal to zero.
	   For both materials the quasiparticle corrections to the OEPx(cLDA) 
	   bandstructure are dispersive. In GaN this effect is slightly more
	   pronounced than in ZnS. The corrections to the LDA
	   bandstructure are less dispersive in GaN, but scatter strongly for 
	   ZnS (see text).}}
\end{figure}

In Fig. \ref{fig:ZnS_GaN_qpeshifts} we have plotted the quasiparticle
corrections of ZnS and GaN as a function of the Kohn-Sham energy. If valence and
conduction bands were shifted uniformly, as this is the case in Silicon for
instance, the circles would form horizontal lines with a jump at the Fermi energy
\cite{Godby/Schlueter/Sham:1988}. Instead the quasiparticle corrections to the
upper valence states of the OEPx(cLDA) ground state decrease linearly with
increasing energy. This leads to a change of the band dispersion in the
quasiparticle bandstructure and we thus speak of \emph{dispersive
quasiparticle shifts}.
The origin for the dispersive nature of the quasiparticle
shift is still unclear and needs to be investigated in the future. The
corrections to the LDA bandstructure of ZnS, however, exhibit two different 
features: one branch shows a similar 
dispersion than in the OEPx(cLDA)+$GW$ case, whereas the corrections to the 
band that hybridises most strongly in energy with the $d$-electrons 
scatter wildy.
This unphysical behaviour is a direct consequence of the inconsistent 
treatment of core-valence exchange in these pseudopotential LDA+$GW$ calculations.

For computational schemes that rely on
core-valence partitioning it is
therefore essential to capture the dominant part of 
core-valence exchange and correlation correctly. Retaining the $s$ and $p$
orbitals of the cation $d$-shell in the core of the pseudopotential will
effectively freeze the core-valence interaction on the level of the density
functional employed, i.e. here LDA or OEPx(cLDA). In the subsequent $GW$
calculation, however, this interaction would be treated by the non-local,
dynamic self-energy. Since the dominant contribution arises from core-valence
exchange, the LDA is not particularly well suited to replace the self-energy. 
As a consequence the perturbation matrix $[\Sigma-v_{xc}]_{nm}$ becomes
non-diagonal and equation \refeq{Eq:qpe} is no longer valid. As a result the
$d$-bands are incorrectly shifted up into the upper valence bands (see left panel of Fig.
\ref{fig:ZnS_BS_b}) and the energy gap opens only by 0.8 eV to 1.60 eV. This
effect was first observed by Rohlfing {\it et al.} \cite{Rohlfing/Pollmann:1995}
(see also line 13 in Tab. \ref{tab:qpe_bg} and 10 in Tab. \ref{tab:dbands}), who
also noted that moving up to second-order perturbation theory in the solution of
equation \refeq{Eq:qpe} gives only marginal improvements.
In order to restore the exchange-interaction,
the $s$ and $p$ electrons would have to be included as valence electrons
in the $GW$ calculation for the Fock part of the self-energy to 
take effect
\cite{Rohlfing/Pollmann:1995,Rohlfing/Pollmann:1998,Luo/Louie:2002} 
(lines 14 and 15 in Tab. \ref{tab:qpe_bg} and 
 lines 11 and 12 in Tab. \ref{tab:dbands}).

\begin{table}
  \caption
  {\small{DFT and quasiparticle band gaps in eV for ZnO, ZnS, CdS, and GaN 
          in the zinc-blende structure sorted in increasing energy from top 
	  to the experimental values. The first column lists the DFT scheme and
	  the second column denoted PP the pseudopotential used. For
	  all-electron calculations this column denotes if the atomic sphere
	  approximation (ASA) or the full potential (FP) was employed. "Conf."
	  refers to the configurations of the (pseudo)atoms: $d$-electrons
	  included ($d$), as described in the previous section, valence only 
	  (no $d$'s), $d$-electrons and their respective shell included
	  ($d$-shell) and all-electron (all $e^-$).	  
          Experimental results are taken from:
                         ZnO \cite{Landolt-Boernstein_1},
			 ZnS \cite{Landolt-Boernstein_2},    
			 CdS \cite{Zahn/Richter:1994},
			 GaN \cite{Ramirez-Flores/etal:1994} 
	  and the OEPx(cLDA) and $GW$ data from:
	  ${}^{a}$Ref. \cite{Zakharov/Louie:1994},
	  ${}^{b}$Ref. \cite{Rubio/etal:1993},
	  ${}^{c}$Ref. \cite{Fitzer/etal:2003},
	  ${}^{d}$Ref. \cite{Staedele/etal:1997},
	  ${}^{e}$Ref. \cite{Aulbur/Staedele/Goerling:2000},
	  ${}^{f}$Ref. \cite{Rohlfing/Pollmann:1995},
	  ${}^{g}$Ref. \cite{Luo/Louie:2002},
	  ${}^{h}$Ref. \cite{Rohlfing/Pollmann:1998},
	  ${}^{i}$Ref. \cite{Usuda/Schilfgaarde:2002}, 
	  ${}^{j}$Ref. \cite{Kotani/Schilfgaarde:2002},
	  ${}^{k}$Ref. \cite{Oshikiri/Aryasetiawan:1999},
	  ${}^{l}$Ref. \cite{Oshikiri/Aryasetiawan:2000}.  
	  The superscript ${}^{w}$ denotes wurtzite structures. In Ref.
	  ${}^{a}$ and ${}^{b}$ a model dielectric function was employed and
	  in Ref. ${}^{e,f,g,h}$ a plasmon pole model was used. }}
  \label{tab:qpe_bg}
   \begin{indented} \item[]
   \begin{tabular}{rllccd{3}d{3}d{3}d{3}}
      \br
      \multicolumn{1}{c}{} &
      \multicolumn{1}{c}{DFT} &
      \multicolumn{1}{c}{PP} &
      \multicolumn{1}{c}{Conf.} &
      \multicolumn{1}{c}{$GW$}&
      \multicolumn{1}{c}{ZnO} &
      \multicolumn{1}{c}{ZnS}&
      \multicolumn{1}{c}{CdS}&
      \multicolumn{1}{c}{GaN} \\ \mr
       1 & LDA       & LDA        & $d$ &	   & 0.51 & 1.76 & 0.81 & 1.65 \\ 
       2 & OEPx(cLDA)& LDA        & $d$ &	   & 1.34 & 2.19 & 1.34 & 2.55 \\ 
       3 & LDA       & LDA        & $d$ & $GW$ & 1.36 & 2.59 & 1.60 & 2.54 \\
       4 & OEPx(cLDA)& LDA        & $d$ & $GW$ & 2.03 & 2.65 & 1.71 & 2.87 \\ 
       5 & OEPx      & OEPx       & $d$ &	   & 2.34 & 2.94 & 1.84 & 2.76 \\ 
       6 & OEPx(cLDA)& OEPx(cLDA) & $d$ &	   & 2.57 & 3.08 & 1.96 & 2.88 \\ 
       7 & OEPx      & OEPx       & $d$ & $GW$ & 3.07 & 3.62 & 2.36 & 3.09 \\
       8 & OEPx(cLDA)& OEPx(cLDA) & $d$ & $GW$ & 3.11 & 3.70 & 2.39 & 3.09 \\ 
       \mr
       9 & \multicolumn{4}{l}{Experiment }
                              & 3.44^{w} & 3.80 & 2.48  & 3.30 \\ \mr 
      10 & LDA        & LDA       & no $d$'s  & $GW$ & & 3.98^{a} & 2.83^{a} 
                                                             & 3.10^{b} \\
      11 & OEPx(cLDA) & OEPx(cLDA)& no $d$'s  &      & & 3.74^{c} &  
                                                             & 3.46^{d} \\
      12 & OEPx(cLDA) & OEPx(cLDA)& no $d$'s  & $GW$ & &          &  
                                                             & 3.49^{e} \\
      13 & LDA        & LDA       & $d$       & $GW$ & & & 1.50^{f} & \\  
      14 & LDA        & LDA       & $d$-shell & $GW$ & & 3.64^{g} & \\  
      15 & LDA        & LDA       & $d$-shell & $GW$ & & 3.50^{h} & 2.45^{h} 
                                                             & 2.88^{h}\\        
      16 & LDA        & FP       & all $e^-$ & $GW$ & 2.44^{w,i} & 3.24^{j}& 
                                                             & 3.03^{w,j}\\ 
      17 & LDA        & ASA      & all $e^-$ & $GW$ & 4.06^{w,l} & 3.97^{k}& 
                                                             & 3.25^{w,l}\\\br
  \end{tabular}
  \end{indented}
\end{table}

In the OEPx approach on the other hand exchange is treated exactly. 
Since we expect dynamic correlation effects to be small for core electrons, 
the dominant contribution is therefore captured by OEPx and we can retain 
the $s$ and $p$ core states in the frozen core of the pseudopotential.
Indeed we observe that the improved description of the
pseudoatoms obtained in OEPx(cLDA) (section \ref{Sec:PPs}) 
translates to the solids
giving band gaps and $d$-electron binding energies systematically closer to 
experiment than in LDA+$GW$.
Including LDA correlation increases the band gaps and $d$-electron binding energies
only slightly  by between 0.1 and 0.2 eV
compared to OEPx (lines 5 and 6 in Tab. \ref{tab:qpe_bg} and \ref{tab:dbands}). 
While the $GW$ formalism proves to be insensitive to this small variation 
for the band gaps (lines 7 and 8 in Tab. \ref{tab:qpe_bg}) the small shift
carries over from OEPx+$GW$ to OEPx(cLDA)+$GW$
for the $d$-electron binding energies
 (lines 7 and 8 in Tab. \ref{tab:dbands}).

For ZnS and CdS OEPx(cLDA) and
OEPx(cLDA)+$GW$ produce essentially the same $d$-electron binding energies. 
Only in ZnO quasiparticle corrections are found to lower
the $d$-states by 1.5 eV compared to OEPx(cLDA), 
further reducing the $p$-$d$ coupling. 
Overall the binding energies obtained with our 
OEPx(cLDA)+$GW$ agree well with other available $GW$ calculations (line 11, 12
and 14 in Tab. \ref{tab:dbands}), but are still about 2 eV at variance with
experiment. Previously Rohlfing \emph{et al}. have devised a $GW$ scheme in 
which plasmon
satellites are included in the Green's function denoted here by SAT
(line 13 in Tab. \ref{tab:dbands}). Although the SAT improves on the 
$d$-electron binding energies the good agreement with experiment for the valence
part of the bandstructure is lost \cite{Rohlfing/Pollmann:1997}. Work towards a
more elaborate theory that provides a description of both the upper valence part
of the bandstructure and the $d$-bands in agreement with photoemssion data
is clearly required in the future.

\begin{table}
  \caption{\label{tab:dbands}{\small 
           $d$-electron binding energies referenced to the top of the valence 
	   band: The layout is the same as in Table \ref{tab:qpe_bg}.
	   Experimental values taken from: 
	       ${}^{a}$Ref. \cite{Weidemann/Becker:1992},
	       ${}^{b}$Ref. \cite{Ley/Shirley:1974},
	       ${}^{c}$Ref. \cite{Ding/etal:1996},
	   and the $GW$ data from:
	       ${}^{d}$Ref. \cite{Rohlfing/Pollmann:1995},
	       ${}^{e}$Ref. \cite{Luo/Louie:2002},
	       ${}^{f}$Ref. \cite{Rohlfing/Pollmann:1998},
	       ${}^{g}$Ref. \cite{Rohlfing/Pollmann:1997},
	       ${}^{h}$Ref. \cite{Usuda/Schilfgaarde:2002},
	       ${}^{i}$Ref. \cite{Kotani/Schilfgaarde:2002},
	       ${}^{j}$Ref. \cite{Oshikiri/Aryasetiawan:2000},
               ${}^{k}$Ref. \cite{Oshikiri/Aryasetiawan:1999}. $SAT$ denotes 
	   $GW$ calculations including plasmon satellites and 
	   the superscript ${}^{w}$ markes studies on the
	   wurtzite structure. 
	   In Ref. ${}^{d,e,f}$ a plasmon pole model was used.%
	       }}

  \begin{indented}
  \item[]
    \begin{tabular}{rllccd{3}d{3}d{3}d{3}}
       \br
      \multicolumn{1}{c}{} &
      \multicolumn{1}{c}{DFT} &
      \multicolumn{1}{c}{PP} &
      \multicolumn{1}{c}{Conf.} &
      \multicolumn{1}{c}{$GW$}&
      \multicolumn{1}{c}{ZnO} &
      \multicolumn{1}{c}{ZnS}&
      \multicolumn{1}{c}{CdS}&
      \multicolumn{1}{c}{GaN} \\ \mr
	1 & LDA	    & LDA	 & $d$ & $GW$ & 4.29 & 4.30 & 6.17 & 13.05 \\
	2 & OEPx(cLDA)& LDA	 & $d$ & $GW$ & 4.98 & 5.02 & 6.40 & 13.58 \\
	3 & OEPx(cLDA)& LDA	 & $d$ &      & 4.36 & 5.33 & 6.54 & 12.75 \\
	4 & LDA	    & LDA	 & $d$ &      & 5.20 & 6.32 & 7.72 & 14.25 \\
	5 & OEPx      & OEPx	 & $d$ &      & 5.12 & 6.91 & 7.57 & 14.85 \\
	6 & OEPx(cLDA)& OEPx(cLDA) & $d$ &      & 5.20 & 7.05 & 7.61 & 15.02 \\
	7 & OEPx      & OEPx	 & $d$ & $GW$ & 6.68 & 6.97 & 7.66 & 16.12 \\
	8 & OEPx(cLDA)& OEPx(cLDA) & $d$ & $GW$ & 6.87 & 7.08 & 7.75 & 16.15 \\
       \mr
        9 & \multicolumn{4}{l}{Experiment }
                              & 9.00^{w,a} & 8.97^{a} & 9.50^{b}  & 17.70^{c} \\ 
          &\multicolumn{4}{l}{}    &            & 9.03^{b} &           & \\ \mr 
       10 & LDA        & LDA	   & $d$       & $GW$ & & & 5.20^{d}
         							   & \\        
       11 & LDA        & LDA	   & $d$-shell & $GW$ & & 7.40^{e} & 
         							   & \\        
       12 & LDA        & LDA	   & $d$-shell & $GW$ & & 6.40^{f} & 8.10^{f} 
         							   & 15.70^{f}\\	
       13 & LDA        & LDA	   & $d$-shell & $SAT$ & & 7.90^{f} & 9.10^{g} 
         							   & 17.30^{f}\\	
       14 & LDA        & FP	   & all $e^-$ & $GW$ & 6.16^{h} & 7.10^{i}  
         					      & 8.20^{i} & 16.40^{w,i}\\ 
       15 & LDA        & ASA	  & all $e^-$ & $GW$ & 5.94^{w,j} & 8.33^{k}& 
                                                             & 17.60^{w,j}\\\br
    \end{tabular}
  \end{indented}
\end{table}

For completeness we have also included previous studies in 
Tab. \ref{tab:qpe_bg} in
which the $d$ electrons were treated as part of the frozen core (lines 10-12).
For reasons given in the previous section these calculations have to be
interpreted cautiously because $p$-$d$ and $d$-$s$ hybridisation is completely 
absent. 

Existing full-potential all-electron LDA+$GW$ calculations (line 16) report band gaps
to within 0.5 to 0.6 eV for GaN and ZnS, but are more at variance for ZnO. 
An underestimation in the RPA
dielectric screening resulting from the LDA ground state was given as a possible
explanation for this discrepancy in Ref. \cite{Usuda/Schilfgaarde:2002}. 
The change in density and
wavefunctions from LDA to OEPx(cLDA) (see previous sections) is
also likely to be an important factor, which requires further analysis. Recently
this conjecture was substantiated by an approximate self-consistent $GW$ 
scheme in which new ground state wavefunctions and a new ground state density 
were calculated from a static but non-local self-energy at every iteration step
\cite{Faleev/Schilfgaarde/Kotani:2004}. Earlier all-electron calculations in the
atomic sphere approximation (ASA) to LMTO were also included in Tab. \ref{tab:qpe_bg} and
\ref{tab:dbands} (lines 15 and 17, respectively). 
The restriction of the potential to a spherical shape inside the atomic spheres
together with the omission of interstitial plane waves in the LMTO leads to an
overestimation of band gaps and $d$-electron binding energies in the ASA.
The seemingly good agreement
with experiment for ZnS and GaN, however, is therefore fortuitous as the 
comparison  with the  more sophisticated full-potential LAPW calculations 
in the lines above illustrates.

To emphasize the importance of consistency concerning the choice of the
exchange-correlation functional in the pseudopotential generation and 
subsequent bulk calculation 
we found it illuminating to include a hybrid calculation in our analysis. 
Line 2 in Tab. 
\ref{tab:qpe_bg} list the values of an OEPx(cLDA) calculation carried
out with LDA pseudopotentials. The gap increases over the LDA values but falls
short of the LDA+$GW$ results. Even more illuminating are the results of the
OEPx(cLDA)+$GW$ hybrid calculation (line 4), which only marginally improve on
the LDA+$GW$ scheme, jeopardizing the good agreement achieved with OEPx(cLDA)
pseudopotenials. This is due to the fact that the self-interaction frozen in the
LDA pseudopotential pushes the $d$-electrons up to the valence bands 
(line 3 in Tab. \ref{tab:dbands}) effectively
closing the $p$-$d$ gap.
The resulting bandstructure looks similar to the LDA+$GW$ bandstucture in 
Fig. \ref{fig:ZnS_BS_b} giving little to no improvement on the LDA results.

Contrary to the LDA, the exchange potential in OEPx
is constructed to best reproduce
the exchange part of the self-energy. OEPx and OEPx(cLDA) can therefore be
regarded as the better and more consistent ground state for a $GW$ excited 
states calculation for
these systems. This conjecture is corroborated by the numerical results 
given in  Tab. \ref{tab:qpe_bg} and \ref{tab:dbands}. 
Summarising the hierachy that has emerged for 
the computational schemes presented here, we find all schemes based on LDA
pseudopotentials at the top of the table and therefore to be the least 
accurate. 
We conclude that using OEPx(cLDA) pseudopotentials in our approach 
is essential. But despite the
considerably opening of the band 
gaps in the OEPx(cLDA) calculations for the four compounds, many-body
perturbation theory in the $GW$ quasiparticle approximation is needed 
to achieve a good 
description of the excitation spectrum and, hence, good
agreement with spectroscopy data.

\subsubsection{Group-III-Nitrides: GaN}
\label{Sec:g3}

\begin{figure}[ht]
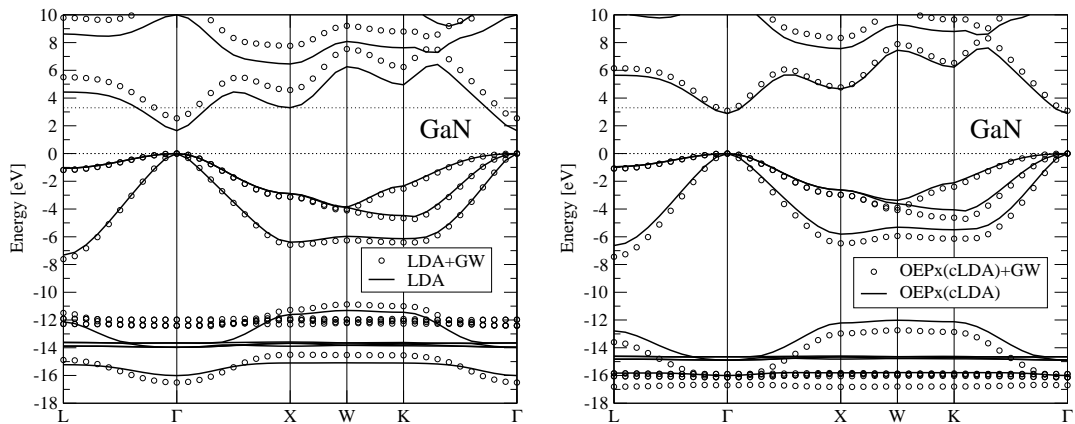

  \begin{center}
    \begin{minipage}[ht]{0.47\linewidth}
      \epsfig{bbllx=39,bblly=71,bburx=512,bbury=453,clip=,
	      file=graphs/GaN_BS_LDA_GW.eps,width=1.0\linewidth}
    \end{minipage}
    \quad
    \begin{minipage}[ht]{0.47\linewidth}
      \epsfig{bbllx=39,bblly=71,bburx=512,bbury=453,clip=,
              file=graphs/GaN_BS_OEPxcLDA_GW.eps,width=1.0\linewidth}
    \end{minipage}
  \end{center}
  \caption{\label{fig:GaN_BS}\small{
           Band structure of GaN in
	   LDA and LDA+$GW$ (panel on the left) compared with
	   OEPx(cLDA) and OEPx(cLDA)+$GW$ (panel on the right). 
	   Consistent pseudopotentials are used.
	   All bandstructures have been aligned at the valence band maximum
	   (dotted line at 0 eV). For reference the experimental band gap is
	   marked by the 2nd dotted line.}}
\end{figure}

As an example for the group-III-nitrides we only consider GaN in this article,
whose bandstructure in our different computational schemes is shown in Fig. 
\ref{fig:GaN_BS}. Values for the band gaps and the $d$-electron binding energies 
\footnote{The center of the $d$-bands for GaN has been obtained by averaging 
over the lowest five bands at the $\Gamma$-point only.} have been
included in Tab. \ref{tab:qpe_bg} and \ref{tab:dbands}. The bandstructure of
GaN clearly reflects the enhanced $s$-$d$ coupling while the $p$-$d$
coupling is greatly reduced compared to the
II-VI semiconductors. As a consequence the upper
valence bands are very similar in all four computational schemes.
The removal of self-interaction in the OEPx(cLDA) based approach (right panel) 
shifts all conduction states
to higher energies compared to the LDA (left panel), 
opening the band gap by 1.2 eV \cite{EXXpaper:2004}. 
The self-energy corrections in the
OEPx(cLDA)+$GW$ calculations, however, are not as pronounced as for
the II-VI semiconductors. This is due to the fact that the
$d$-electrons are much lower in energy and more strongly localised on the 
cation and hence do not contribute as much to the correlation part of the 
self-energy as in the case of the II-VI compounds. 
The Ga $d$ and N $s$ bands in our OEPx(cLDA)+$GW$ are in very good agreement 
with $GW$ calculations by Rohlfing \emph{et al}. \cite{Rohlfing/Pollmann:1998}
with small differences only at the $\Gamma$-point.

The quasiparticle shifts to the
LDA ground state are dispersionless for the lower conduction bands,
as Fig. \ref{fig:ZnS_GaN_qpeshifts} illustrates, but scatter
in an energy window of $\pm$0.2 eV around a value of -0.6 eV for the upper
valence states. The self-energy correction to the OEPx(cLDA) ground state on the
other hand decreases almost linearly with increasing energy and is thus much 
more dispersive than for the LDA ground state
(compare also the left and right panel of Fig. \ref{fig:GaN_BS}). 

It remains to be added that all statements made in the 
previous section about the comparison between the OEPx and
OEPx(cLDA) results and those of the respective $GW$ calculations also hold 
true for GaN.
For the OEPx(cLDA) hybrid approach with LDA pseudopotentials we observe two
different regimes. Due to 
the reduced $p$-$d$ coupling in GaN, the removal
of the self-interaction among the valence electrons leads to small improvements
for the band gap in the OEPx(cLDA) compared to the LDA results
(lines 2 and 4 in Tab. \ref{tab:qpe_bg}).
The Ga 3$d$ binding energies and the dispersion of the semicore bands, however,
are comparable to those in the LDA (lines 2 and 4 in Tab. \ref{tab:dbands}). 

\subsubsection{Electronic Self-Energy}
\label{Sec:Sigma}

We will close our investigation into the electronic structure of the four
compounds presented here with an analysis of the electronic self-energy.
Since we apply $GW$ non-self consistently as a perturbation to the Kohn-Sham
bandstructure ($\{\epsilon_{n\vk}^{\rm DFT}\}$) without diagonalising the
quasiparticle Hamiltonian the quasiparticle energies 
($\{\epsilon_{n\vk}^{qp}\}$) comprise four 
contributions\footnote{For clarity we have repeated equation \refeq{Eq:qpe}
here.}
\begin{equation}
\label{Eq:qpe_2}
  \epsilon_{n\vk}^{qp}=\epsilon_{n\vk}^{\rm DFT}+\bracket{\phi_{n\vk}}
                       {\Sigma_{xc}^{GW}(\epsilon_{n\vk}^{qp})-v_{xc}-\Delta\mu}
		       {\phi_{n\vk}} \quad .
\end{equation}
A closer look at equation \refeq{Eq:qpe_2} reveals that the
quasiparticle energies can change because a) the Kohn-Sham bandstructure to which
the self-energy corrections are applied changes b) the ingredients (Kohn-Sham
energies and wavefunctions) for
the self-energy operator have changed and/or c) because the
wavefunctions used for the evaluation of the matrix elements in equation 
\refeq{Eq:qpe_2} differ from one ground state to another.

To ascertain which of these three factors dominantes
we have compiled the individual contributions\footnote{
For the remainder of the discussion we have introduced $\expv{A}$ as a short 
hand notation for the matrix 
elements $\bracket{\phi_{n\vk}}{A}{\phi_{n\vk}}$. 
The alignment constant $\Delta \mu$ has been absorbed into the matrix
elements of $v_{xc}$.} 
that enter
equation \refeq{Eq:qpe_2} in Table \ref{tab:sigcontrib} for three
selected states of ZnS and GaN.
In order to isolate the influence of the Kohn-Sham bandstructure it is conducive
to separated the perturbation $\expv{\Sigma-v_{xc}}$ 
into a static $\expv{\Sigma_x-v_{xc}}$ and a dynamic\footnote{The distinction
between \emph{static} and \emph{dynamic} is made here purely on the grounds of
distinguishing between explicitly energy dependent (dynamic) and energy
independent quantities (static). It is not to be understood in the quantum
chemical sense as the difference between Hartree-Fock and dynamic correlation
methods.}  
$\expv{\Sigma_{c}(\epsilon^{qp})}$ contribution evaluated at the 
converged quasiparticle energy. The static part depends only on
the wavefunctions of the occupied states (see equation \refeq{Eq:def:sigma_x})
and the wavefunction of the bandstructure state for which the quasiparticle
energy is evaluated, whereas the correlation part of the self-energy is
additionally dependent on the wavefunctions of all unoccupied states and
the Kohn-Sham eigenvalues via the dynamic
polarisability (see equation \refeq{Eq:G_0} to \refeq{Eq:def:sigma_c}).

\begin{table}
  \caption{\label{tab:sigcontrib}{\small 
           Contributions to the quasiparticle energy shift as
	   applied via equation \refeq{Eq:qpe_2} for ZnS and GaN and three
	   different representative states: the conduction band minimum (CBM),
	   the valence band maximum (VBM) and the highest (lowest) $d$-electron
	   state for Zns (GaN) at the $\Gamma$-point ($d$-state). The columns
	   list the DFT eigenvalues ($\epsilon^{\rm DFT}$) for the respective 
	   ground state (GS) calculation, the matrix elements of the 
	   exchange part ($\expv{\Sigma_x}$, equation \refeq{Eq:def:sigma_x}) and the 
	   correlation part ($\expv{\Sigma_c}$, equation \refeq{Eq:def:sigma_c})
	   of the self-energy, 
	   the matrix elements of the exchange-correlation potential 
	   ($\expv{v_{xc}}$, including the alignment constant $\Delta \mu$ 
	   (equation \refeq{Eq:qpe})) and the static contribution 
	   ($\expv{\Sigma_x-v_{xc}}$). (All energies are given in eV).
	       }}
  \begin{indented}
  \item[]
    \begin{tabular}{lcld{3}d{3}d{3}d{3}d{3}c}
       \br
      \multicolumn{1}{c}{} &
      \multicolumn{1}{c}{State} &
      \multicolumn{1}{c}{GS} &
      \multicolumn{1}{c}{$\epsilon^{\rm DFT}$} &
      \multicolumn{1}{c}{$\expv{\Sigma_x}$} &
      \multicolumn{1}{c}{$\expv{v_{xc}}$} &
      \multicolumn{1}{c}{$\expv{\Sigma_c}$} &
      \multicolumn{2}{c}{$\expv{\Sigma_x-v_{xc}}$} \\ \mr
      \multirow{6}{0.7cm}{ZnS}	   
        &\multirow{2}{0.9cm}{CBM} 
	         & LDA        & 1.75 &  -7.14 & -11.82 & -4.33 & 4.68 \\
	       & & OEPx(cLDA) & 3.08 &  -6.56 & -11.26 & -4.39 & 4.69 \\\ns  
	       & & \crule{7} \\ 
        &\multirow{2}{0.9cm}{VBM}  
                 & LDA        & 0.00 & -17.64 & -15.60 & 1.62 & -2.04 \\
	       & & OEPx(cLDA) & 0.00 & -17.60 & -15.73 & 1.57 & -1.87 \\\ns 
	       & & \crule{7} \\ 
        &\multirow{2}{0.9cm}{$d$-state}  
                 & LDA        & -6.62 & -28.19 & -24.74 & 4.95 & -3.46 \\
	       & & OEPx(cLDA) & -7.29 & -29.01 & -23.29 & 5.27 & -5.71 \\ \mr
      \multirow{6}{0.7cm}{GaN}	   
        &\multirow{2}{0.9cm}{CBM} 
	         & LDA        & 1.65 & -10.05 & -14.76 & -4.27 & 4.72 \\
	       & & OEPx(cLDA) & 2.88 &  -9.23 & -13.61 & -4.28 & 4.38 \\\ns  
	       & & \crule{7} \\ 
        &\multirow{2}{0.9cm}{VBM} 
                 & LDA        & 0.00 & -21.15 & -18.40 &  2.32 & -2.76 \\
	       & & OEPx(cLDA) & 0.00 & -21.45 & -19.03 &  2.32 & -2.43 \\ 
	       & & \crule{7} \\ 
        &\multirow{2}{0.9cm}{$d$-state}  
                 & LDA        & -13.97 & -31.77 & -27.40 & 5.19 & -4.37 \\
	       & & OEPx(cLDA) & -14.89 & -32.46 & -25.61 & 5.50 & -6.85 \\ 
      \br
    \end{tabular}
  \end{indented}
\end{table}

Focusing first on the conduction band minimum (CBM) and the 
valence band maximum (VBM) of ZnS and GaN we observe that the correlation contributions
of the self-energy agree to within 0.05 eV for the respective states 
and thus prove to be insensitive to changes in the ground state from LDA to OEPx(cLDA).
The same is true for the static contribution, which changes up to only 0.3 eV,
although both $\expv{\Sigma_x}$ and $\expv{v_{xc}}$ taken individually
exhibit larger variations. This pattern is transferable to the other upper
valence and conduction states and to ZnO and CdS. From this we conclude that the
perturbation $\expv{\Sigma-v_{xc}}$ is small for this part of the
energy spectrum and largely insensitive to variations 
in the exchange-correlation functional from LDA to exact-exchange 
in the II-VI compounds. In GaN the difference in the quasiparticle shifts
between the two $GW$ calculations is more pronounced 
(see also Fig. \ref{fig:ZnS_GaN_qpeshifts}) but with $\sim$0.7 eV 
still amounts to only half of the band gap difference between LDA and
OEPx(cLDA).
The strong improvement we have reported in the previous sections
for the OEPx(cLDA)+$GW$ approach is therefore, to a large extend, due to the 
changes in the
Kohn-Sham bandstructure. 

For the $d$-bands, however, the situation is drastically different. The
correlation contribution is still very similar with differences around 0.3 eV,
but the static contribution differs vastly by up to 2.5 eV for the case of GaN.
In addition the perturbation operator is no longer diagonal in the LDA+$GW$
calculations \cite{Rohlfing/Pollmann:1995,Luo/Louie:2002} and the $d$-bands are
pushed upwards into the upper valence states in the II-VI compounds 
\cite{Rohlfing/Pollmann:1995} (see Fig. \ref{fig:ZnS_BS_b}). This is a direct
consequence of the inconsistent description of core-valence exchange for the
cation in the LDA+$GW$ approach. 
We therefore expect that in full-shell LDA+$GW$ calculations the
static contribution for the $d$-electrons becomes comparable in size 
to the values for our OEPx(cLDA)+$GW$ calculations in which core-valence exchange is treated much
more consistently through the OEPx pseudopotentials. At the same time this
modified exchange self-energy will affect the perturbation operator for the 
valence states much more, which then leads to a lowering of the occupied 
states relative to the conduction states and thus to the opening of the 
band gap that is observed in 
full-shell \cite{Rohlfing/Pollmann:1995,Rohlfing/Pollmann:1998,
                  Luo/Louie:2002,Zakharov/Louie:1994}
and all-electron 
\cite{Kotani/Schilfgaarde:2002,Usuda/Schilfgaarde:2002,
      Faleev/Schilfgaarde/Kotani:2004}
LDA+$GW$ calculations. A numerical verification of
this hypothesis will be given elsewhere.
\section{Conclusions}

We have reported a first combined OEPx and $GW$ study investigating
the effects of semicore states on the electronic structure of selected II-VI 
compounds and group-III-nitrides.
The removal of the self-interaction in the OEPx(cLDA) approach leads to a
stronger localisation of the cation $d$-electrons compared to the LDA and thus 
to a reduced $p$-$d$ hybridisation. 
As a result both the gap between $d$ and upper
valence bands as well as between valence and conduction bands
opens up in the OEPx(cLDA) approach. 
Switching to the $GW$ picture of interacting quasiparticles 
we obtain band gaps in very 
good agreement with photo-electron spectroscopy, 
provided of course we start from an OEPx or OEPx(cLDA) ground state. 

The self-energy correction shifts the conduction and upper valence bands almost
rigidly when starting from an LDA calculation, whereas in the case of OEPx(cLDA)
ground states the corrections are dispersive.
Dispersive self-energy shifts are relatively uncommon and not frequently
reported in the literature; a point that requires further investigation in the
future.

We find that the $d$-electron binding energies in our OEPx(cLDA)+$GW$ approach 
are in good agreement with those of previously reported $GW$ calculations  
but are still at variance
with experiment. The fact that the erroneous energetical position of the 
$d$-electrons actually produces quasiparticle energies for the 
upper valence and conduction bands in agreement with experiment
indicates that the $GW$ approximation might not suffice
to fully describe the strongly localised cation $d$-electrons. A more
elaborate theory incorporating electron-hole and/or
vertex corrections might be necessary in order to obtain a bandstructure
consistent with photo-electron spectroscopy over a larger energy window.

Consistency in the choice of the pseudopotential was found to be paramount.
Moreover, employing OEPx(cLDA) pseudopotentials allows us to remove 
the $s$ and $p$ electrons of the respective $d$-shell from the calculation by
freezing them in the core of the pseudopotential.
This reduces the required plane-wave cutoff and thus the computational costs 
of our calculations tremendously. It remains to be verified, however, how the
pseudoisation of the atomic wavefunctions affects the exchange-integrals in the
OEPx approach. This will be the subject of future studies.

Furthermore we have  alluded to the formal connection between the 
exchange potential in 
the OEPx formalism  and the exchange part of the self-energy. 
Our numerical results confirm the
hypothesis that for the class of materials presented here the DFT 
exact-exchange ground state constitutes a much better starting point
for $GW$ bandstructure calculations.
In light of this we like to conclude with the outlook that 
the Sham-Schl\"uter equation \refeq{Eq:def:ShScheq} 
offers a formal but heuristic perspective to include quasiparticle 
correlation into the OEP ground state and thus to go beyond OEPx+$GW$.

\section*{Acknowledgements}
We like to ackowledge Matthias Wahn and Philipp Eggert for many stimulating
discussions and Martin Fuchs for his invaluable expertise on 
pseudopotentials.
Moreover, we are indebted to A. Majewski and P. Vogl for making their 
pseudopotential generator available to us.
This work was in part funded by the Volkswagen Stiftung/Germany and
the EU's 6th Framework Programme through the NANOQUANTA Network of 
Excellence (NMP4-CT-2004-500198).
\clearpage
\section*{References}

\bibliographystyle{phpf}
\bibliography{./OEPx-3d-v5}

\end{document}